\documentclass[aps,prl,reprint,superscriptaddress,longbibliography]{revtex4-1}
\usepackage[english]{babel}
\usepackage{siunitx}  
\usepackage{graphicx} 
\usepackage{miller}   
\usepackage[percent]{overpic} 
\usepackage{pifont}
\usepackage{xcolor} 
\usepackage[utf8]{inputenc}
\DeclareSIUnit\rydberg{Ry}
\usepackage{mathtools,amssymb}
\DeclareSIUnit\sq{\ensuremath{\Box}}

\begin{document}

\title{Electronic properties of bismuth nanostructures}

\author{Christian \surname{König}}
\affiliation{Tyndall National Institute, University College Cork, Lee Maltings, Cork T12 R5CP, Ireland}
\author{James C. \surname{Greer}} \email[Corresponding author:~]{Jim.Greer@nottingham.edu.cn}
\affiliation{Nottingham Ningbo New Material Institute and Department of Electrical and Electronic Engineering, University of Nottingham Ningbo China, 199 Taikang East Road, Ningbo, 315100, China}
\author{Stephen \surname{Fahy}}
\affiliation{Tyndall National Institute, University College Cork, Lee Maltings, Cork T12 R5CP, Ireland}
\affiliation{Department of Physics, University College Cork, College Road, Cork T12 K8AF, Ireland}

\date{\today}

\begin{abstract}
The passivation of thin Bi\hkl(1 1 1) films with hydrogen and
oxide capping layers is investigated from first principles.
Considering termination-related changes of the crystal structure,
we show how the bands and density of states are affected.
In the context of the much discussed semimetal-to-semiconductor
transition and the band topology of the bulk material,
we consider the effects of confinement in the whole Brillouin
zone and go beyond standard density functional theory by including
many-body interactions via the G$_0$W$_0$ approximation.
The conductivity of unterminated films is calculated via the
Boltzmann transport equation using the simple constant relaxation
time approximation and compared to experimental observations
that have suggested a two-channel model.
This article got published in Physical Review B: https://journals.aps.org/prb/abstract/10.1103/PhysRevB.104.045432
\end{abstract}

\maketitle

\section{Introduction}
\begin{figure*}
\center
\includegraphics{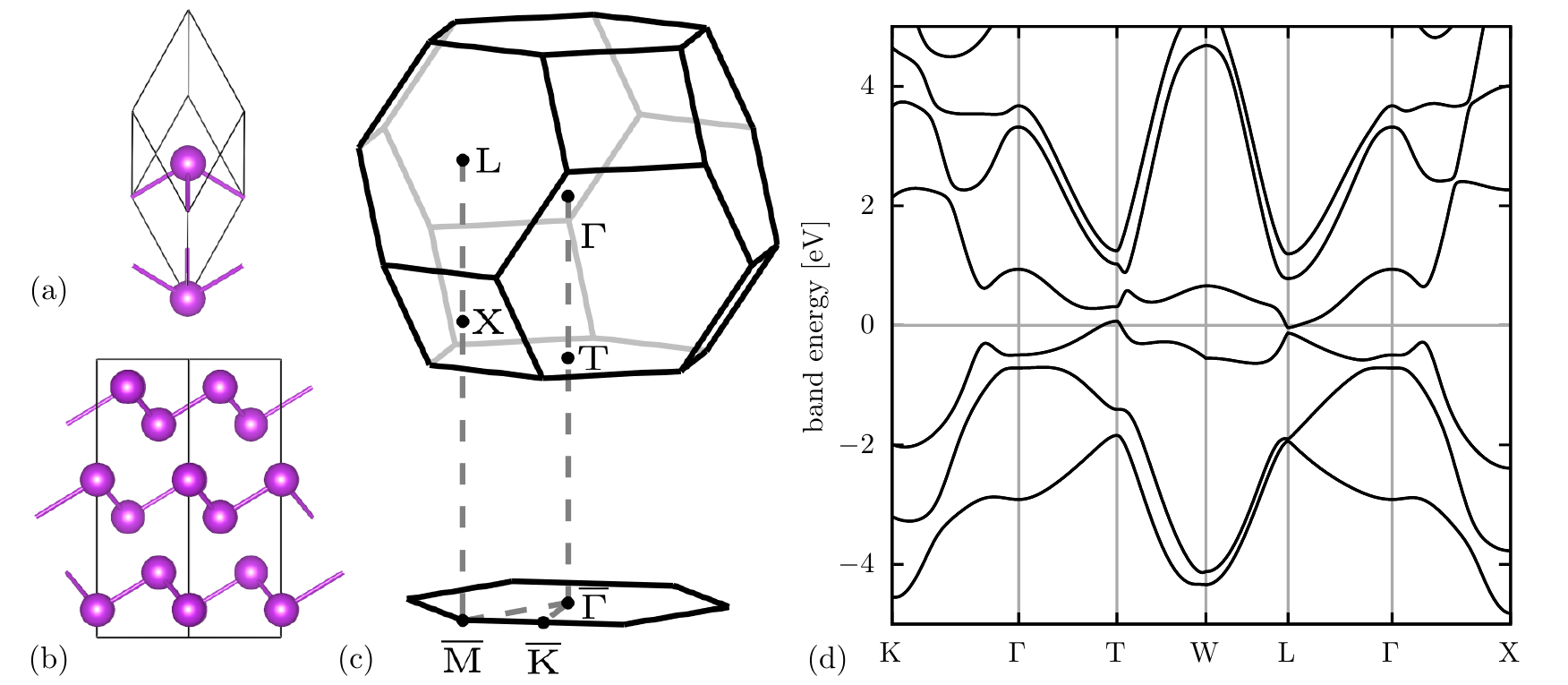}
\caption{ \label{fig:cell}
  (a) Rhombohedral Bi unit cell. The second basis atom is slightly
  displaced from the center of the cell. Therefore, bismuth forms
  bilayers which are clearly visible in (b) where we show a side
  view of the hexagonal supercell (black lines, contains only six atoms).
  The bilayers lie perpendicular to the \hkl[1 1 1] direction.
  (c) Brillouin zone of the unit cell with the main high
  symmetry points and their projection onto the two-dimensional
  hexagonal Brillouin zone of the Bi\hkl(1 1 1) surface.
  (d) Band structure of the bulk as calculated with density functional
  theory.   All bands are doubly degenerate due to inversion and time
  reversal symmetry (Kramers degeneracy theorem).
  For more details on the structure we refer to Ref.~\cite{hofmann2006}.
}
\end{figure*}

The electronic properties of the semimetal bismuth have been subject
to intense investigation. The bulk material has a low
density of states at the Fermi level which originates from the indirect
overlap in the range of \SIrange{30}{40}{\milli\electronvolt}
\cite{aguilera2015,edelman1975,isaacson1969b,dinger1973,smith1964}
between the valence band at T and the conduction band at L.
The direct band gap is located at the L point and even smaller with a value of
only about \SIrange{10}{15}{\milli\electronvolt}
\cite{aguilera2015,maltz1970,vecchi1974,isaacson1969b,brown1963,smith1964}.
These delicate features make the material interesting to study but
on the other hand also challenging to describe with standard
electronic structure methods. Figure \ref{fig:cell} shows the
crystal structure and electronic structure of the bulk.

Bismuth nanostructures offer a means to probe and manipulate material
properties and thus are interesting both from physics and application points
of view. Experimental studies have in particular investigated films
ranging from the micrometer to the nanometer range. The small effective
masses of the charge carriers at the Fermi level lead to confinement
effects which can increase the energy difference between the valence and
conduction bands (see e.g.~Refs.~\cite{hirahara2006,hirahara2007b,ishida2016, ito2016,cantele2017})
and a semimetal-to-semiconductor transition in the nanometer thickness regime has been anticipated and studied
by many groups, see, e.g., Refs.~\cite{lutskii1965,ogrin1966,sandomirskii1967,hoffman1993,rogacheva2008}.
Although the transition into the semiconducting regime offers great potential
for applications \cite{gity2017,gity2018}, it is still subject to an ongoing
debate and the growing importance of surface states with increasing surface-to-volume
ratio in very thin films has to be considered. The surface state conductivity of Bi\hkl(1 1 1)
films has been found to be particularly large
and in recent years various
transport measurements \cite{hirahara2007,fei2010,xiao2012,zhu2016,hirahara2018,zhu2018,kroeger2018},
and magnetotransport measurements \cite{luekermann2013,aitani2014,abdelbarey2020}
suggested that the conductivity of thin films is composed of two channels,
i.e.~a semiconducting and a metallic contribution, ascribed to the film
interior and surface.
The passivation of the surface states surprisingly has not been studied to the same extent.

It has been noted, e.g., in Refs.~\cite{aitani2014,zhu2016,kroeger2018,abdelbarey2020},
that although the two-channel model is quite successful in describing the measurements,
the surface and core regions can not be considered to be completely decoupled. In
particular, Ishida \cite{ishida2016} showed that the surface states at the $\overline{\Gamma}$
and $\overline{\text{M}}$ point extend considerably into the core of the film. Accordingly, the effect of
quantum confinement is most pronounced in these regions. The surface states of the
films also play a significant role in the discussion of the topological properties
of bismuth and their ma\-ni\-fes\-ta\-tion in two-dimensional films affected by
quantum confinement, see, e.g., Ref.~\cite{chang2019}. Three-dimensional topologically
nontrivial materials have surface states, which are immune to passivation \cite{hasan2010}.

Various methods have been used in order to obtain the band structure of bismuth although not
all of them are easily transferable to nanostructures. The tight-binding model developed by
Liu and Allen \cite{liu1995} is commonly used in the literature and aims to reproduce the bulk
features around the Fermi level. To do so, interactions up to the third-nearest neighbors had
to be included.
As pointed out in Refs.~\cite{fukui2007,ohtsubo2013,ohtsubo2016}, minimal changes in the model
produce basically the same features but induce a topological phase transition.
In addition to the uncertainty related to the model (or any other electronic structure method
for that matter), another disadvantage of the tight-binding model is that surfaces, which
naturally occur in nano\-struc\-tures are not well described. Therefore, additional surface hopping
terms like in Ref.~\cite{saito2016} may have to be introduced. Specifically, a simple truncation
of the bulk model leads to a crossing of the surface states of Bi\hkl(1 1 1) \cite{ohtsubo2016},
which are important in nanostructures, e.g., for the overall density of states in thin films.
Furthermore, tight-binding models are not easily transferable and predictions for different
types of surface terminations are not possible, making the method not ideal for nanostructures.

Density functional theory (DFT) is a ground state theory and well suited to predict the
structure of materials from first principles. However, the electronic states are not
accurate by construction,
with the exception of the highest valence state, which is the ionization energy \cite{aryasetiawan1998,almbladh1985}.
The many-body problem is only approximated with the exchange-correlation functional and
the states are only meant to reproduce the true charge density for the system (rather than quasiparticle energies) \cite{aryasetiawan1998}.
Nevertheless, DFT is often used for calculations of the band structure.
As it turns out, the results often are reasonable with exception of the notorious
underestimation of the band gap. Therefore, with respect to the Bi tight-binding model,
DFT suffers from similar issues since an artificial inversion of the small L gap might
occur, with consequences for the topology of the bulk material.

The problems of the methods described above can be minimized by means of the GW method, which corrects
the state energies as calculated with DFT by replacing the exchange-correlation
functional with the calculation of the self-energy, see, e.g., Ref.~\cite{aryasetiawan1998}.
Thus, the interaction and screening between the electrons is more accurately described,
leading in general to more reliable results and a much better agreement between the calculated
band gap and its experimental value \cite{aryasetiawan1998}.
Naturally, the computational cost increases accordingly.

In the following we use density functional theory with complementary G$_0$W$_0$ calculations in order
to investigate if the surface states in thin Bi\hkl(1 1 1) films can be passivated.
We furthermore discuss the effects of confinement in the films as calculated with both
methods as well as the consequences for the band gap and conductivity.


\section{Computational Details}
The plane-wave code \textsc{Quantum Espresso} \cite{giannozzi2009,giannozzi2017}
was used for all density functional theory calculations.
The generalized gradient approximation (GGA) as formulated by Perdew,
Burke, and Ernzerhof \cite{perdew1996} was employed using fully-relativistic
and norm-conserving SG15 pseudopotentials \cite{hamann2013,schlipf2015,scherpelz2016}.
We note that spin-orbit coup\-ling is essential for a good description of the
electronic structure of bismuth and is taken into account in every step of
our calculations, including the G$_0$W$_0$ correction of the state energies
for which we used \textsc{Yambo} \cite{marini2009,sangalli2019}.

In the following we discuss the properties of thin bismuth films with different geometries.
To start with, we show the effect of confinement on the surface states of Bi\hkl(1 1 1)
films as calculated with DFT. In order to allow for a comparison with the bulk and previous
calculations, we used the experimental lattice parameters from Ref.~\cite{schiferl1969}
measured at roughly \SI{300}{\kelvin}, which are also used in Ref.~\cite{aguilera2015}.
A kinetic-energy cutoff of $\SI{50}{\rydberg}$ and a kinetic charge-density cutoff of
$\SI{200}{\rydberg}$ proved to be sufficient to very accurately obtain the state energies
on a $6 \times 6 \times 1$ $\mathbf{k}$-point grid.
The vacuum was set to \SI{20}{\angstrom}, which is more than sufficient to suppress
interactions between the per\-io\-dic images of the slab.
A bulk calculation in a hexagonal supercell was done for the projected bulk states.
The results are compared to those of a G$_0$W$_0$ calculation.

In addition to that, the effect of surface termination on the structures previously described in Ref.~\cite{koenig2019}
with regards to a passivation of the surface states is discussed. These structures
were optimized (under constraints) so that the remaining forces do not exceed \SI{5e-2}{\electronvolt/\angstrom}.
While accurate forces require a kinetic-energy cutoff of up to \SI{100}{\rydberg},
the state energies converge faster, which reduces the computational cost,
in particular regarding the G$_0$W$_0$ calculations.
Therefore, in general, for all structures
a cutoff of \SI{60}{\rydberg} or more was used for the calculation of DFT states,
which can serve as input for G$_0$W$_0$.

We use the random phase and plasmon pole approximation implemented in \textsc{Yambo}.
Notably, G$_0$W$_0$ requires more $\mathbf{k}$-points than DFT to converge.
In order to ensure reliable results, our calculations use up to
$30 \times 30 \times 1$ $\mathbf{k}$-points. For even better accuracy
of the $\overline{\text{M}}$ gap we employed an extrapolation scheme.
Further details are given below.
Furthermore, a Coulomb cutoff \cite{rozzi2006} has to be used in \textsc{Yambo}
to completely avoid the interaction between the slab and its periodic images.
The particular implementation requires even more vacuum than
the DFT calculations and
the overall cell height has to be more than twice the thickness of the slab.
The combination of both effects, and of course the total number of electrons
in the system, limits the maximum film thickness for which the many-body
corrections can be calculated.

Since \textsc{Yambo} only determines the quasiparticle energies on a uniform grid
of $\mathbf{k}$-points, an interpolation method has to be used to plot the band structure.
An interpolation to a fine grid in $\mathbf{k}$-space is also useful in order to
calculate the conductivity. Thus, we interpolate the bands with \textsc{BoltzTraP2}
\cite{madsen2006,madsen2018} via a smooth Fourier interpolation scheme
\cite{euwema1969,shankland1971,koelling1986,pickett1988} which exactly reproduces
the input data. The density of states and conductivity are subsequently
calculated by a bespoke code developed in-house.


\section{Surface states on the B\MakeLowercase{i}\hkl(1 1 1) surface}
\begin{figure*} 
\center
\includegraphics[width=\textwidth]{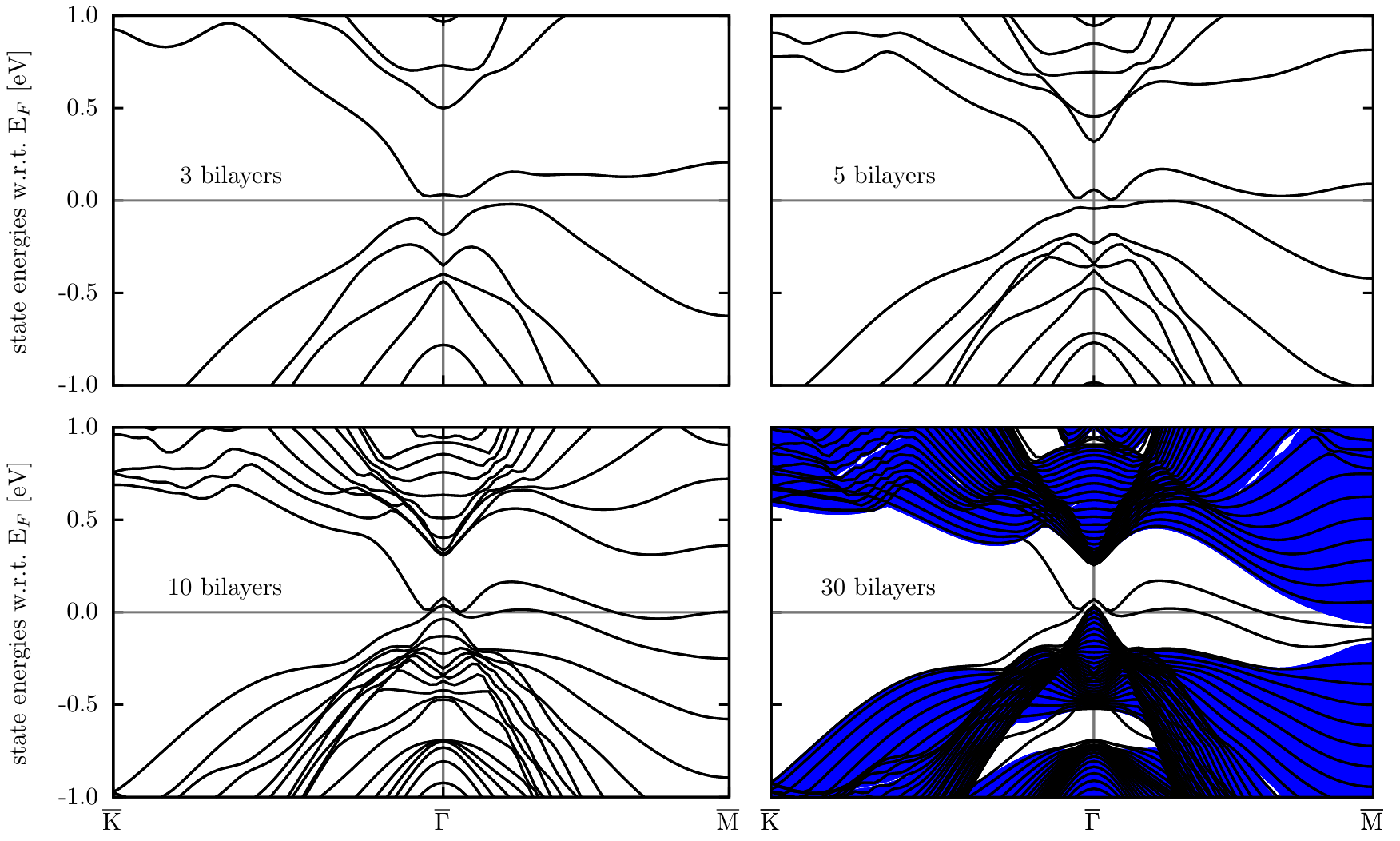}
\caption{ \label{fig:surfaceStatesAndProjection}
  DFT band structures for films with 3, 5, 10, and 30 bilayers
  (i.e., \SI{1.0}{\nano\meter}, \SI{1.7}{\nano\meter}, \SI{3.7}{\nano\meter}, and \SI{11.6}{\nano\meter}) thickness.
  The slabs were cut from the bulk with experimental geometry;
  no further relaxation was allowed.
  The number of quantum-well states grows with increasing thickness,
  approaching the bulk limit. By the same token, the splitting of the
  valence and conduction bands is strong only in thin films and gradually
  decreases. In line with Ref.~\cite{ishida2016}, mainly $\overline{\Gamma}$ and
  in particular $\overline{\text{M}}$ are affected by confinement.
  For 30 bilayers we also show the projected bulk band structure.
  Here the effect of confinement at $\overline{\Gamma}$ is small, so that the
  band edges give a clear energy reference where we can align the
  projected band structure to. Since the surface states do not
  connect to the valence and conduction bands at $\overline{\text{M}}$ any more,
  it follows that the DFT calculation predicts a trivial band topology
  for bulk bismuth, see, e.g., Fig.~3 (c) in Ref.~\cite{ohtsubo2013}.
  At $\overline{\Gamma}$, the surface states are degenerate
  but do not connect to the bulk valence bands.
  This is in contrast to Refs.~\cite{ito2016,yao2016,ohtsubo2016}, presumably
  due to the lack of a clear energy reference as only the Fermi level is used for the relative alignment of the bands to the projected bulk states.
}
\end{figure*}

Figure \ref{fig:surfaceStatesAndProjection} shows the band structure
of thin unterminated Bi\hkl(1 1 1) slabs with a thickness between
$3$ and $30$ bilayers.
By comparison with the projected band structure of the bulk (shaded region in the figure),
the two surface bands can be identified at the Fermi level as they appear
in the projected bulk gap.
In fact, each of the bands is degenerate with another band and
the two states correspond to the two surfaces of the slab.
Only in a semi-infinite bulk, there is no second surface 
and therefore no degeneracy so that
two spin-split bands remain on the surface.
The states in a slab are subject to quantum confinement perpendicular
to the surface so that the energy splitting between the valence and
conduction states increases with respect to the bulk.
Since the direct band overlap of Bi is small, a
semimetal-to-semiconductor transition has been expected even for
relatively thick films (between \SIrange{23}{32}{\nano\meter}) 
\cite{lutskii1965,sandomirskii1967,hoffman1993,hofmann2006,rogacheva2008}.
Nevertheless, the experimental observations made are not without
ambiguity and the transition is still subject to debate.

It is clear from Fig.~\ref{fig:surfaceStatesAndProjection} that
the effect of confinement is most prominent at $\overline{\text{K}}$, $\overline{\Gamma}$, and
particularly $\overline{\text{M}}$. The states between $\overline{\Gamma}$ and $\overline{\text{M}}$ remain relatively close to the Fermi level.
For details we refer to Ref.~\cite{ishida2016}.
Overall, the expected onset of a semiconducting phase with a large band gap
is not in agreement with the predictions of density functional theory.
We can only see a very small gap for three bilayers or about \SI{1}{\nano\meter} thickness, which
is also sensitive to further relaxations away from the experimental bulk lattice.
In fact, the expectations formulated above rely on an idealized picture of quantum confinement as a perturbation to the bulk.
The fact that broken bonds may give rise to metallic states at the surface is not considered.
In other words, it is assumed that the surface states are removed from the band structure by passivation.
This passivation, e.g., with a native oxide, may be required in Bi films to open the band gap (see also Ref.~\cite{hirahara2007}).
In the following we investigate if a hydrogen, hydroxyl or Bi$_2$O$_3$ termination as presented in Ref.~\cite{koenig2019} can achieve passivation of the surface bands.

The discussion will be divided into two sections.
First of all, we start by using the experimental bulk geometry for a three bilayer thick film.
Thus we can clearly isolate the effect of the surface termination without any side effects,
e.g., further relaxation of the films. Each side of the slab will be terminated with hydrogen or
hydroxyl, which is allowed to relax on the surface. For the oxide termination, a five bilayer slab
is used where the upper and lower bilayers are oxidized and relaxed.
Afterwards, we consider further relaxation of the structures and apply strain equivalent to
lattice matched growth of Bi\hkl(1 1 1) on a Si\hkl(1 1 1)--$7 \times 7$ substrate.

\begin{figure*} 
\center
\includegraphics{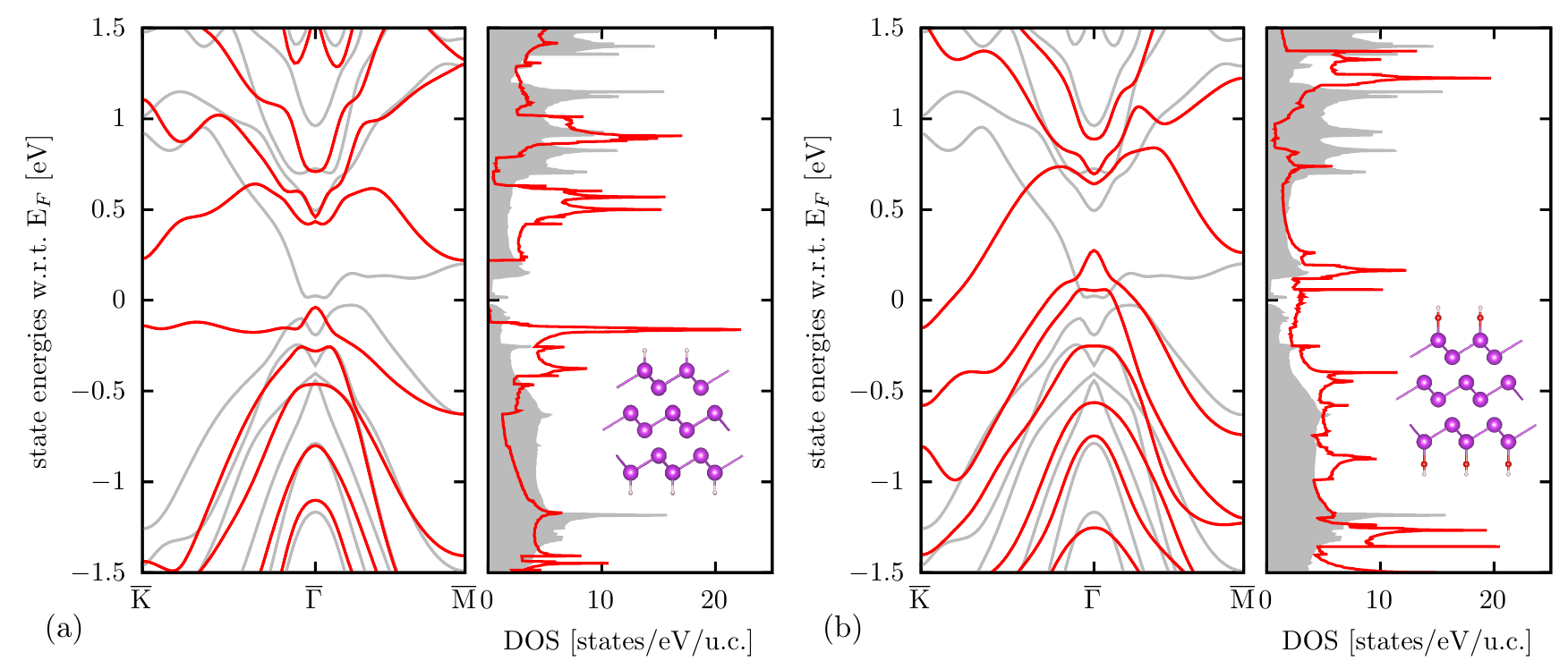}
\caption{ \label{fig:simpleTerminationHandOH}
  DFT:
  (a) Band structure and DOS of a hydrogen terminated slab. For simplicity, the
  slab was kept in the experimental geometry of the bulk and only the H atoms were relaxed.
  We also show the data for the unterminated slab in gray.
  Clearly, the points $\overline{\Gamma}$ and $\overline{\text{K}}$ are affected by the surface termination. At $\overline{\Gamma}$,
  the conduction bands are pushed to higher energies, which increases the overall distance
  between the occupied and unoccupied states in the Brillouin zone. The previously very
  large local gap at $\overline{\text{K}}$ is reduced while the $\overline{\text{M}}$ point is only marginally affected.
  In this artificial but controlled setting, we observe a passivation of the surface, which
  results in a band gap of \SI{259}{\milli\electronvolt}.
  (b) Results for the hydroxyl termination. In comparison to (a), the main difference
  is the $\overline{\text{K}}$ point where the lowest conduction band (effectively one of the two surface bands)
  drops below the Fermi level so that the system becomes even more metallic. This is clearly
  reflected in the DOS.
}
\end{figure*}

Figure \ref{fig:simpleTerminationHandOH} shows the bands and density of states (DOS) of
the -H and -OH terminated structures and compares the results to the unterminated slab.
We can see that the hydrogen termination does reduce the DOS at the Fermi level
by pushing the lowest conduction band states at $\overline{\Gamma}$ to higher binding energies.
In turn, the splitting at $\overline{\text{K}}$ is reduced -- but not enough to close the gap.
As we will show subsequently, further relaxation of the structure will close
the gap again.
The hydroxyl termination has a similar effect on the band structure, but the
changes at $\overline{\text{K}}$ are even more pronounced. The first conduction band
crosses the Fermi level and closes the global gap due to an overlap with
the valence bands at $\overline{\Gamma}$.

\begin{figure*} 
\center
\includegraphics{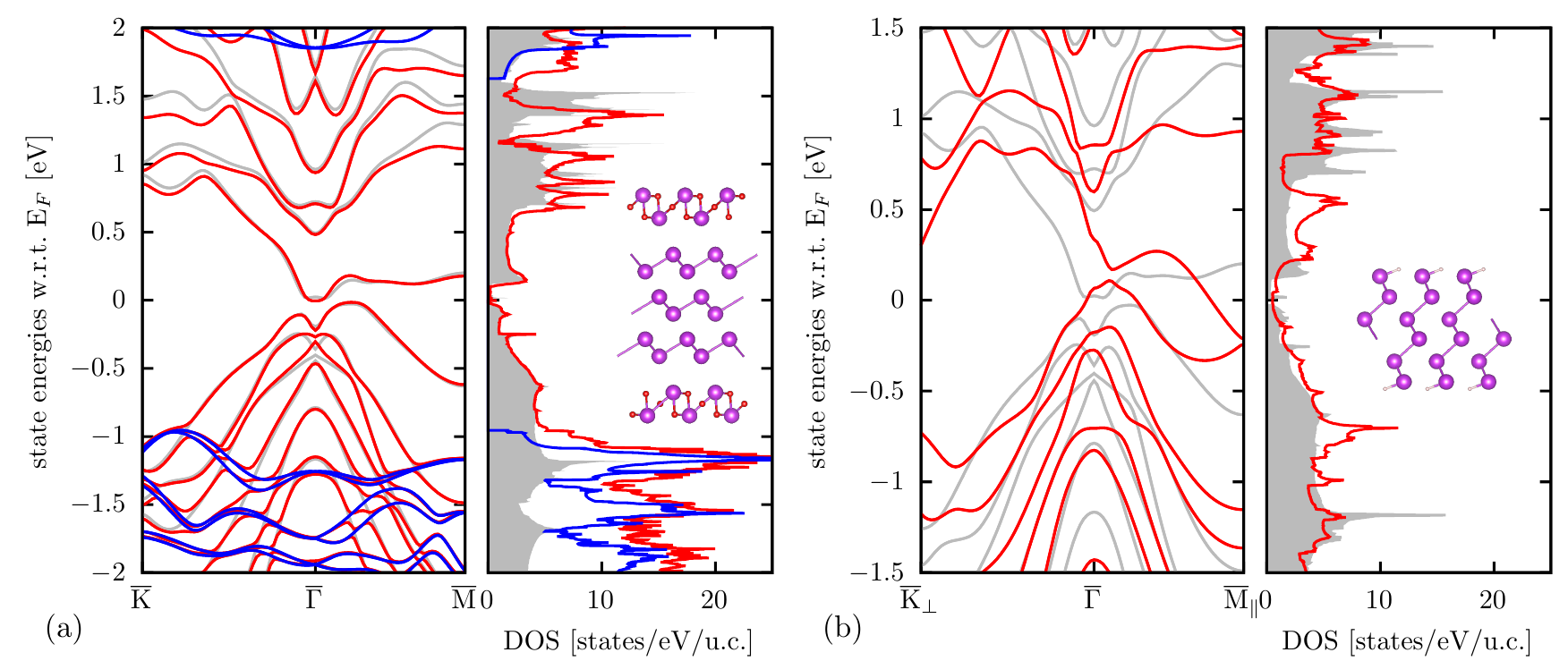}
\caption{ \label{fig:oxideAndRelaxedH}
  DFT:
  (a) Band structure and density of states for the oxide terminated slab in red, the
  unterminated slab in gray, and the corresponding isolated oxide layers in blue.
  While the Bi core region was kept in the experimental bulk geometry, the oxide was relaxed on the surface
  of the film and shows very little interaction with the core. In particular, the surface
  bands at the Fermi level are not affected and the band structure is basically a superposition
  of those of the sub-systems.
  Thus, the oxide termination has no passivating effect on the surface states.
  The bands of the relaxed hydrogen terminated structure are shown in (b) where the
  in-plane lattice constant was kept fixed. The reorientation of the film
  has a significant impact on the electronic structure.
  The C$_3$ symmetry is lost, so that the $\overline{\text{M}}$ points (and $\overline{\text{K}}$ points) are no longer equivalent.
  Thus, we have chosen a path, which is perpendicular and parallel to the characteristic direction
  of the system, given by the orientation of the surface bonds.
  The valence and conduction bands overlap in the direction of the Bi-H bonds and
  therefore the system remains semimetallic.
}
\center
\includegraphics{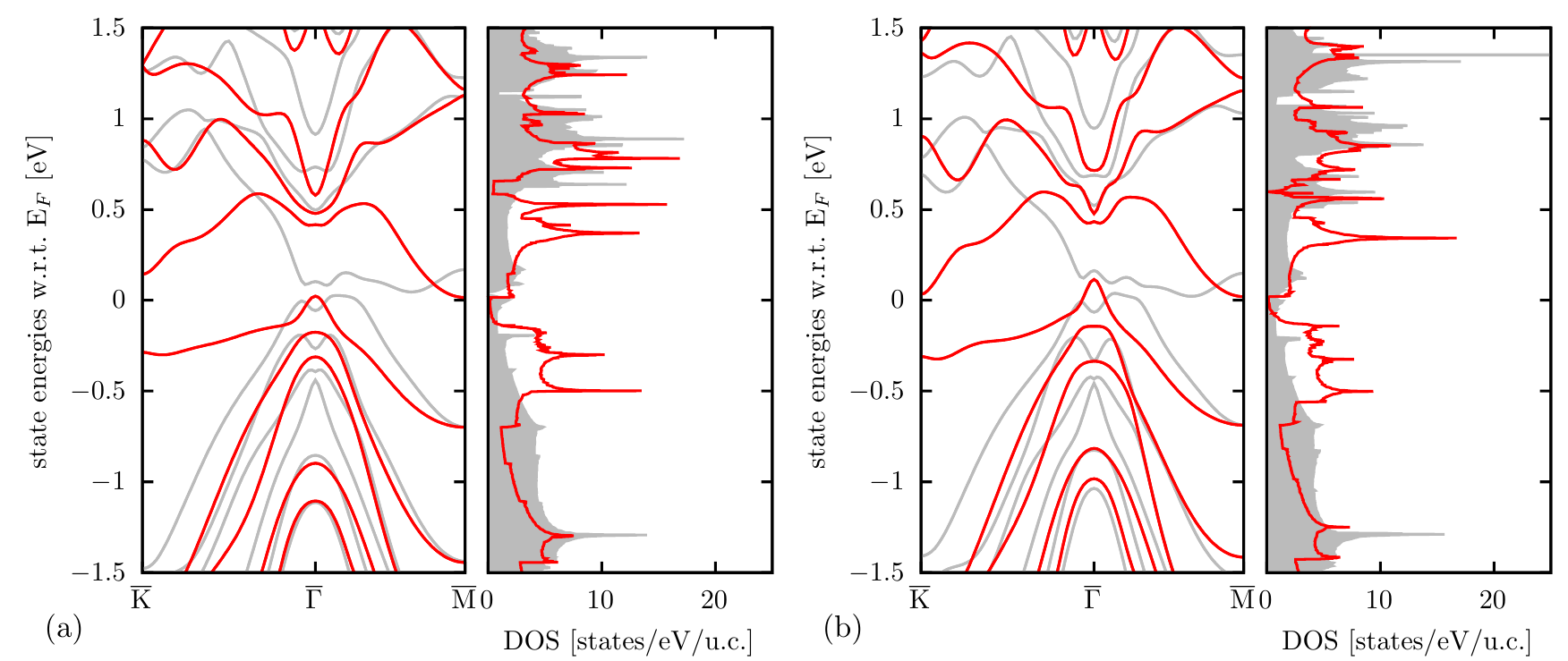}
\caption{ \label{fig:relaxationAndStrainEffects}
  DFT:
  The figure shows the effects of relaxation and strain on three bilayer thick films. The results for the unterminated slabs are shown in gray while the red curves correspond to those with termination. The slab in (a) was cut from
  a bulk lattice where the out-of-plane lattice constant $c_h$ and atomic positions were optimized via a minimum energy fit (GGA). Effectively the
  inter-bilayer distance increased while $a_h$ was not changed with respect to the
  previous calculations. As a consequence, the T-L overlap increases in the unterminated slab
  (dip in the surface state between $\overline{\Gamma}$ and $\overline{\text{M}}$) although the splitting due to
  confinement is much stronger at $\overline{\text{M}}$. However, compared to the previous crystal
  structure, the overlap at $\overline{\text{M}}$ also increases if H is attached to the surface and
  the DOS remains semimetallic even with termination.
  Panel (b) corresponds to a thin film on a Si\hkl(1 1 1)--$7 \times 7$ substrate,
  i.e., \SI{1.3}{\percent} in-plane compressive strain. In order to avoid issues
  with the interlayer distance, we deduced $c_h$ by assuming constant volume.
  The atomic positions in the bulk were also optimized via a fit.
  We observe an effect similar to (a) with increasing indirect band overlap
  and an almost identical semimetallic density of states at the Fermi level.
  However, in contrast to (a), this not due to the deficiencies of DFT
  regarding the equilibrium structure.
}
\end{figure*}

The oxide termination in Fig.~\ref{fig:oxideAndRelaxedH} (a) on the other hand
basically does not have any significant effect on the surface states. The equilibrium position
of the oxide layers is further away from the Bi core than another Bi bilayer would be.
Figure \ref{fig:oxideAndRelaxedH} (a) shows the band structures of the full terminated slab, the core Bi region,
and the isolated oxide layers. Clearly, the
surface bands are affected only slightly and there appears not to be enough chemical
interaction with the oxide to push the states away from the Fermi level.
The band dispersion of the valence bands below \SI{-1}{\electronvolt} is to a good
approximation just a superposition of the bands of the two sub-systems. Since GGA
underestimates the bond strengths, we also investigated how the electronic structure
is affected if the oxide layers are closer to the surface of the Bi slab.
Keeping those Bi atoms of the oxide layers, which are closest to the slab in the
position of a new bilayer, the interaction is enhanced slightly. However, rather than
passivating the slab, the states at $\overline{\Gamma}$ are affected such that the small
band gap is closed completely.

We note that the band gap of our oxide layers (approximately \SI{2.6}{\electronvolt})
is in good agreement with experimental values for $\alpha$-Bi$_2$O$_3$ which is the
most common and stable oxide phase \cite{leontie2000,leontie2001,leontie2002,walsh2006}.
The native oxide on the films in Ref.~\cite{gity2018} was found to be Bi$_2$O$_3$.
Five different phases of bulk Bi$_2$O$_3$ are known and the native oxide layer may,
as is shown in Refs.~\cite{leontie2000,leontie2001,leontie2002} for Bi$_2$O$_3$ films,
consist of a mixture of these phases (or even amorphous regions and regions with different
stoichiometry, also depending on the preparation conditions).
Values of \SI{2.15}{\electronvolt} and \SI{2.60}{\electronvolt} for the indirect (thermal and optical) band gap in $\alpha$-Bi$_2$O$_3$ and \SI{2.91}{\electronvolt} for the direct (optical) gap were reported in Ref.~\cite{gobrecht1969}.
However, there is considerable variation to be expected (\SIrange{2.3}{3.3}{\electronvolt}
for the optical gap in the oxide films, \SI{2.3}{\electronvolt} for $\alpha$-Bi$_2$O$_3$,
and about \SI{2}{\electronvolt} for amorphous films in Ref.~\cite{leontie2002}).
The structure of the native oxide and the layered oxide model may not coincide with the bulk phases, so that the electronic structure is modified. Confinement may also further change the band gap in the thin oxide layers. 
Nevertheless, overall the model appears to describe the oxide layers reasonably well
\footnote{The band gap is wide enough so that the coupling of the oxide to the Bi states at the Fermi level is small.}.

After this simplified investigation, there are two additional factors, which have to be considered for the
realistic description of the films.
First of all, the slabs might relax further, in particular perpendicular to the substrate.
Chemical interaction with the goal of passivating the surface will have an impact on
the overall structure.
This has been discussed in Ref.~\cite{koenig2019}
for hydrogen and hydroxyl termination, which changed the orientation of the whole film for sufficiently small thickness.
A stronger covalent bonding between film and oxide than observed here for our model can be expected to lead to a
change in the orientation of the crystal close to the surface.
Furthermore, lattice strain due to the growth on a mismatched substrate such as Si\hkl(1 1 1)--$7 \times 7$ has to be considered.

In any case, we require a relaxation of the full structure with respect to the experimental bulk lattice.
We note that the weak forces between the layers in combination with the high number of atomic coordinates
makes it difficult for common relaxation algorithms to find the true global energy minimum.
To start with, Fig.~\ref{fig:oxideAndRelaxedH} (b) shows the hydrogen terminated film after relaxation
where the in-plane lattice constant was kept fixed at
\SI{4.546}{\angstrom}. We observe the expected
reorientation of the film induced by the covalent bonds at the
surface, which has a significant impact on the electronic structure of the film. Since the rotational
symmetry of the previously hexagonal surface is lost, we have chosen those $\overline{\text{K}}$ and $\overline{\text{M}}$ points for the band
structure plot, which are perpendicular and parallel with the in-plane orientation of the Bi-H bond.
The overlap between the valence and conduction band has increased significantly due to the changes at the
$\overline{\text{M}}$ point and the film has become even more metallic.
Therefore, it is not realistic to assume that a hydrogen termination would lead to the passivation
seen in the unrelaxed geometry shown in Fig.~\ref{fig:simpleTerminationHandOH} (a).

There are some further intricacies of the electronic structure related to
the equilibrium geometry of the film. Figure \ref{fig:relaxationAndStrainEffects} (a) shows the band
structure and DOS of a bismuth film with the same in-plane lattice constant as the previously
discussed films.
The remaining parameters that determine the crystal structure were
optimized for a minimum bulk energy.
Since GGA underestimates the bond strengths, the interlayer distance increases.
As a consequence, as we have discussed previously in Ref.~\cite{koenig2021a},
the indirect band overlap between T and L increases \footnote{At the same time the
material is pushed towards the topologically nontrivial state (decreasing L gap).}.
These high-symmetry points of the bulk project to the $\overline{\Gamma}$ and $\overline{\text{M}}$ point of the slab, see Fig.~\ref{fig:cell} (c).
Indeed, compared to Fig.~\ref{fig:simpleTerminationHandOH} (a), the surface state of the unterminated slab (drawn in gray) between these two points drops
in Fig.~\ref{fig:relaxationAndStrainEffects} (a) to lower energies and closes the gap of the unterminated slab.
Directly at $\overline{\text{M}}$, the effect of confinement is much stronger and pushes the bands apart.
Hydrogen was then used to passivate the surface and only the H atoms were relaxed.
The corresponding band structure of the terminated slab is drawn in red.
Notably, in contrast to the case of experimental lattice geometry, the first
conduction state at $\overline{\text{M}}$ now drops in energy, leading to a semimetallic density of states.

\begin{figure*} 
\center
\includegraphics{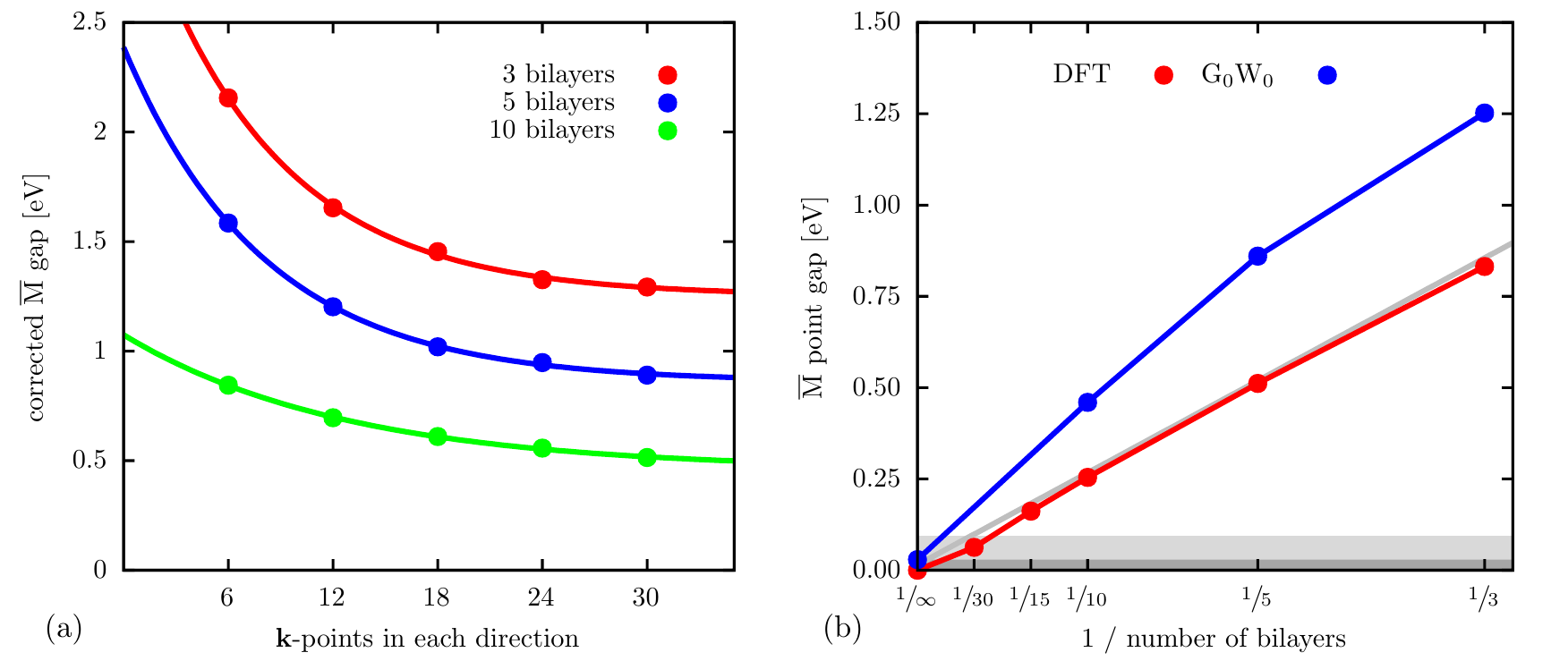}
\caption{ \label{fig:MGap}
  (a) The direct G$_0$W$_0$ $\overline{\text{M}}$ gap was fitted as a function of the number of $\mathbf{k}$-points
  in each direction of the Brillouin zone.
  A decaying exponential was used in order to obtain the converged value.
  The results are shown in (b) as a function of the reciprocal slab thickness
  and compared with the DFT values.
  The light and dark gray areas indicate the L gap of the bulk
  as calculated in DFT or G$_0$W$_0$, respectively.
  In DFT, the band topology of bismuth is trivial, which is confirmed
  by the fact that the value of the $\overline{\text{M}}$ gap for 30 bilayers is already
  smaller than the bulk L gap.
  In the semi-infinite limit, the surface states still exist but
  are degenerate as shown in Refs.~\cite{ishida2016,chang2019,jin2020} and as a result the gap does vanish.
  In the case of G$_0$W$_0$, we have shown previously \cite{koenig2021a}
  that Bi is topologically nontrivial (using the same crystal structure and methods).
  Thus, one of the surface bands connects to the valence band while
  the other surface band connects to the conduction band in the semi-infinite limit.
  The same interpolation was done for the experiments in Refs.~\cite{ito2016,ito2020}
  and the respective data combined with that from Ref.~\cite{hirahara2007b} is shown as a gray line.
  The data contains only films with at least $7$ bilayers so that the line is extrapolated to thinner films.
  Surprisingly, the experimental results match very well with the DFT results
  whereas G$_0$W$_0$ overestimates the $\overline{\text{M}}$ gap.
}
\end{figure*}

Next, we consider the strain, which a Si\hkl(1 1 1)--$7 \times 7$ substrate exerts on
a thin Bi film. The strain in films of this thickness has been reported
to be \SI{-1.3}{\percent} so that the bismuth film matches with the silicon
substrate with six unit cells of the hexagonal lattice
\cite{nagao2004,kammler2005,nagao2005,yaginuma2007}.
In order to avoid any issues related to the bilayer distance as predicted
by the GGA, we apply the strain to the in-plane lattice parameter $a_h$ and
adjust the out-of-plane pa\-ra\-me\-ter $c_h$ accordingly so that the cell volume
is preserved; the coordinates of the basis atoms were optimized for the bulk.
The results are shown in Fig.~\ref{fig:relaxationAndStrainEffects} (b).
Again, the gray bands correspond to the unterminated slab for which the
results are very similar to Fig.~\ref{fig:relaxationAndStrainEffects} (a).
Clearly, the compressive strain in the plane increases the overlap of the
valence and conduction bands in line with the discussion in
Ref.~\cite{koenig2021a} for the bulk.
Introducing bonding hydrogens to both sides of the film (bands drawn in red) does not passivate
the surface since the overlap between the valence and conduction bands persists.

We note again that the relaxation and strain effects also affect the unterminated
three bilayer slab in Fig.~\ref{fig:surfaceStatesAndProjection} insofar as the small band gap, which is obtained with
experimental bulk geometry, closes.


\begin{figure*} 
\center
\includegraphics{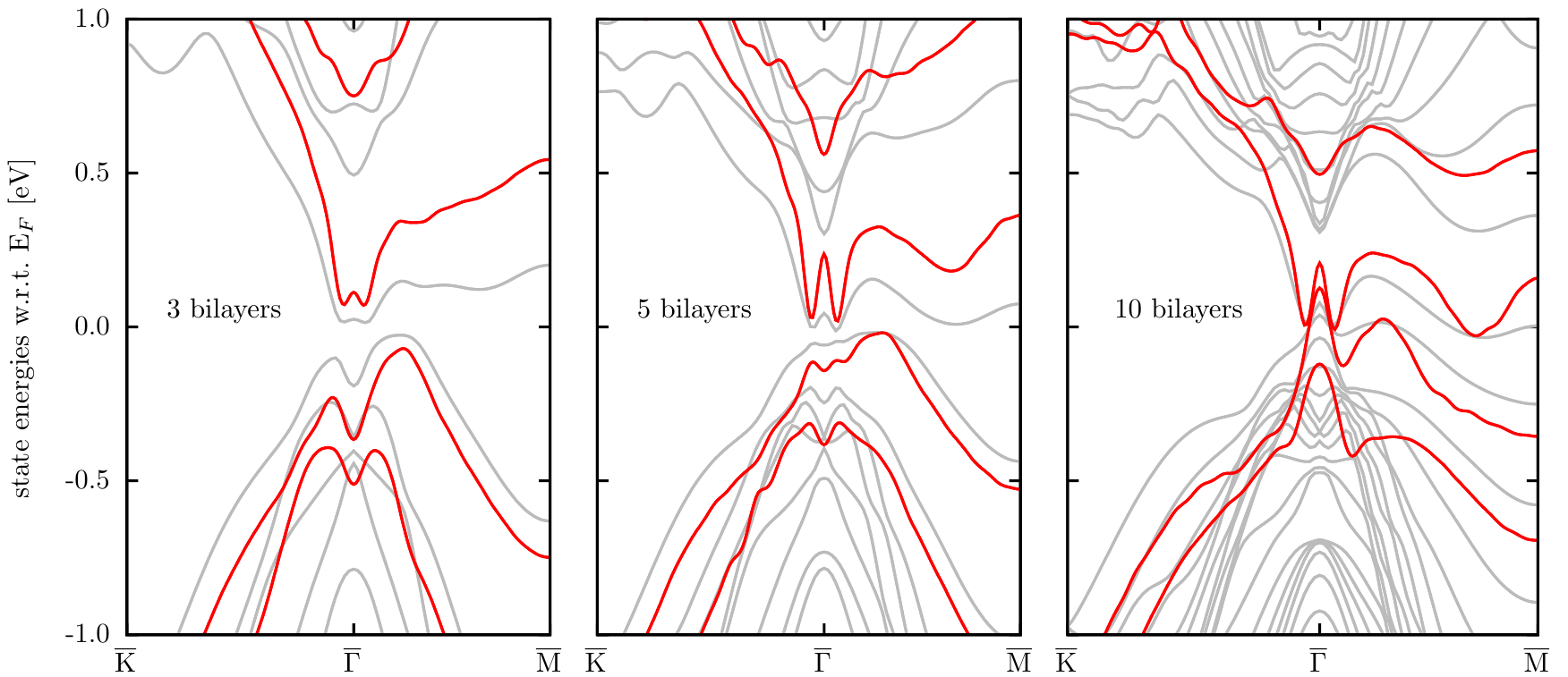}
\caption{ \label{fig:G0W0Bands}
  Band structure of 3, 5, and 10 bilayer thick Bi\hkl(1 1 1) films in bulk-like geometry with (red)
  and without (gray) the many-body corrections.
  Only the first four valence and conduction bands of the G$_0$W$_0$ results
  were interpolated and plotted for each slab thickness.
  The interpolation with \textsc{BoltzTraP2} ensures that there are no fitting errors,
  in particular at $\overline{\text{K}}$, $\overline{\Gamma}$, and $\overline{\text{M}}$, which however causes
  some Gibbs oscillations on the path between the calculated points.
  The Fermi level is at \SI{0}{\electronvolt}.
}
\end{figure*}

\section{Confinement: DFT and G$_0$W$_0$}
Based on the observations above, the surface states can not realistically be
expected to be removed in our DFT calculations \footnote{
Only in the artificial situation
where hydrogen is attached to an otherwise unrelaxed slab, the surface states are
sufficiently deformed to open a small band gap, but not entirely removed.
}.
In-plane compressive strain or a relaxation towards increasing bilayer
distance is further detrimental to passivation since the overlap of the
valence and conduction bands increases.
Therefore, the reported semiconducting behavior of thin films
raises the question if density functional theory achieves an
adequate description of the confined system.
Since the $\overline{\text{M}}$ gap is the most sensitive to confinement and also
accessible in experiment via angle-resolved photoemission
spectroscopy (ARPES), this quantity is a reasonable observable
to check.

We calculated the $\overline{\text{M}}$ gap for films with a thickness between
3 and 30 bilayers, again using the experimental bulk lattice.
The results are shown in Fig.~\ref{fig:MGap}.
There we also compare the DFT results to those of G$_0$W$_0$ calculations, which for many
materials improve the electronic structure, in particular regarding
the energy gaps. As shown in Fig.~\ref{fig:MGap} (a), a high number
of $\mathbf{k}$-points is required in the two-dimensional G$_0$W$_0$
calculations and we use an extra\-po\-la\-tion scheme in order to
obtain well converged results.
The limiting factors that determine the maximum film thickness
for which calculations are possible are discussed above in the
methods section.

Figure \ref{fig:MGap} (b) shows the same analysis that
was applied to the experimental results in Ref.~\cite{hirahara2007b}
and Refs.~\cite{ito2016,ito2020}
where the measured $\overline{\text{M}}$ gaps were plotted as a function of the
reciprocal film thickness and used for an extra\-po\-la\-tion
to the semi-infinite limit. Since the gap is increased in
thin films due to confinement, the relative error is small and
the extra\-po\-la\-tion is thus more reliable than a direct
measurement of the small splitting of the surface bands on the semi-infinite surface,
which at the maximum might amount to a value of the order of the bulk L gap.
We note that in comparison to the films calculated here, those
used in experiment are typically thicker, since these samples
are easier to grow with high quality.

A linear dependence of the $\overline{\text{M}}$ gap as a function of the reciprocal
slab thickness has been found in Ref.~\cite{ito2016} over a range
of $14$ to $202$ bilayers
which is explained in Ref.~\cite{ito2020} by the Dirac dispersion
of the bulk in the direction normal to the film surface;
the $k_\perp$ of the standing waves are determined by
the thickness and by taking into account the phase shifts at the surface
and at the interface to the substrate.
Our DFT and G$_0$W$_0$ results agree with this observation
reasonably well even for a thickness of only a few bilayers.
We note that a gap, which is a linear function of the reciprocal
thickness has also been taken as an indication of a parabolic
confinement potential, e.g., in Ref.~\cite{kroeger2018}.

Surprisingly, the DFT results for the $\overline{\text{M}}$ gap are
already in good agreement with the experimental data while G$_0$W$_0$
overestimates the gap substantially.
From this perspective, the DFT description of the surface states
in thin Bi\hkl(1 1 1) films, affected by confinement, appears to
be very reasonable.
However, the discrepancy of the many-body calculations requires some
clarification, particularly since G$_0$W$_0$ is known to improve
the description of the bulk bands \cite{aguilera2015,koenig2021a}.

\begin{figure*} 
\center
\includegraphics{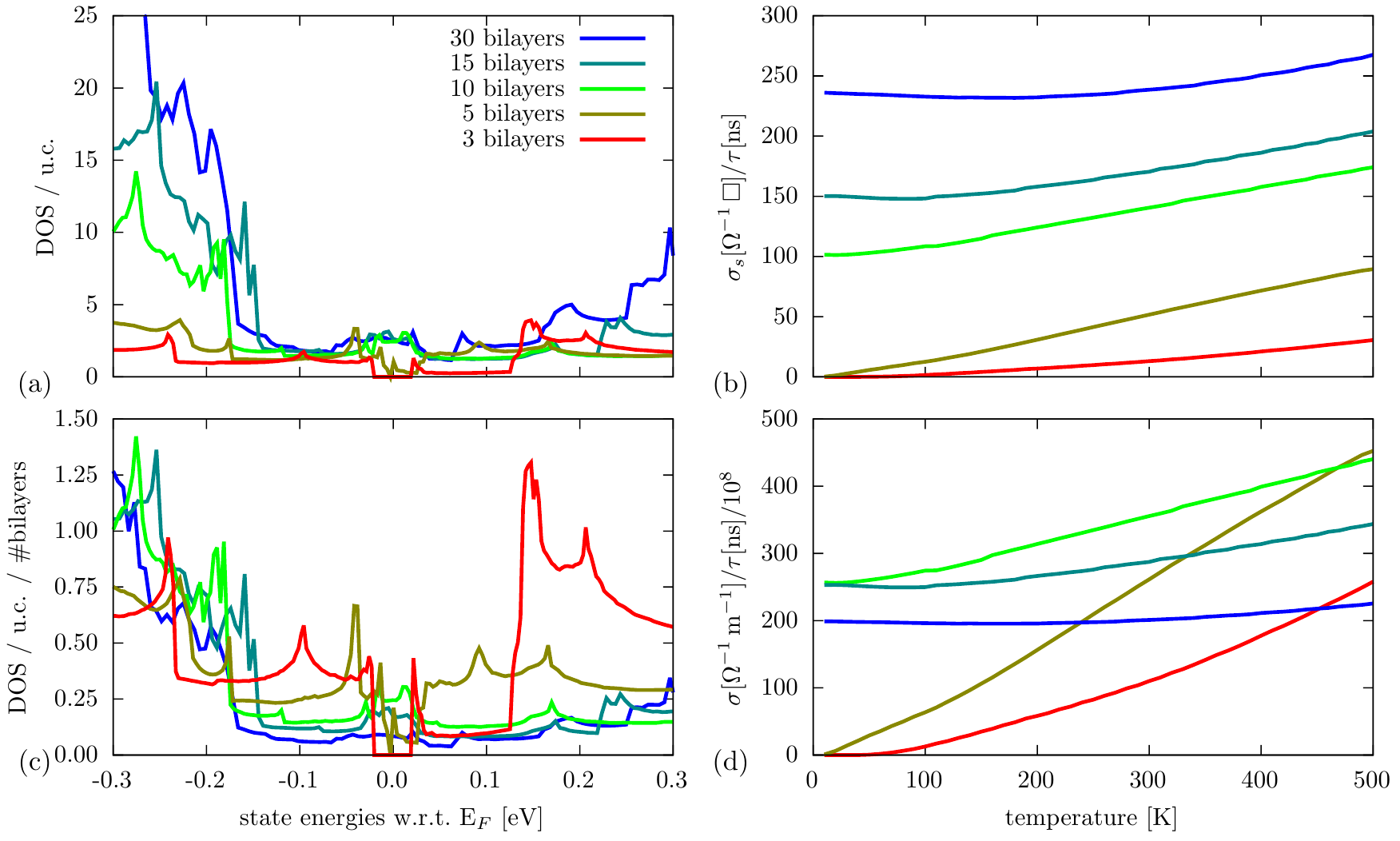}
\caption{ \label{fig:dosAndSigma}
  (a) DFT density of states per two-dimensional unit cell for slabs with different thickness.
  The band gap of the three bilayer thick film closes for five bilayers.
  Thicker films all have about the same amount of states near the Fermi level which is the surface state contribution. (b) shows the corresponding sheet conductivity (depending on the constant relaxation time $\tau$).
  Slabs with a thickness of ten or more bilayers already conduct for very low temperatures
  since their surface states cross the Fermi level and are occupied even without thermal excitation.
  Panel (c) shows the density of states normalized by the slab thickness.
  The number of valence states per energy in the unit cell is converging quickly towards the bulk limit with increasing thickness.
  The conductivity in (d) corresponds to (b) divided by the slab thickness.
  In each case the arithmetic mean of the two in-plane directions is shown.
}
\end{figure*}


\section{Full many-body band structure}
Knowing only the size of the $\overline{\text{M}}$ gap is not sufficient in the context of
a semimetal-to-semiconductor transition and for the interpretation of
the conductivity measurements in the literature. Rather, the calculation
of many $\mathbf{k}$-points in the Brillouin zone equivalent to the full
band structure is required. We evaluated the quasiparticle energies
on the full $30 \times 30$ $\mathbf{k}$-point grid. Figure \ref{fig:G0W0Bands}
shows the interpolated band structures for $3$, $5$, and $10$ bilayer thick
slabs in bulk-like geometry and compares to the DFT bands.

As we have observed previously, the splitting of the surface states at $\overline{\text{M}}$ is
enhanced with respect to DFT and the same is true for $\overline{\text{K}}$. The most interesting
however is the behavior in the middle of the Brillouin zone near the $\overline{\Gamma}$ point, where the DFT valence and conduction bands
are very close. The many-body corrections split the
bands further apart so that for three bilayers a substantial band gap can be observed.
Thus, G$_0$W$_0$ suggests that quantum confinement alone is sufficient to open a small band gap of about \SI{140}{\milli\electronvolt}
in the very thin film and no passivation of the surface states is required. We note however that this gap is smaller
than commonly expected for a total thickness of only \SI{1.0}{\nano\meter}.
Furthermore, the gap already closes for less than ten bilayers, i.e., \SI{3.7}{\nano\meter} thickness.

Unfortunately, there is not the same amount of data in the literature for other symmetry points than $\overline{\text{M}}$.
Based on our observations, the $\overline{\Gamma}$ point of the films and its contribution
to the conductivity would be particularly interesting
and could offer an explanation for the observed semiconducting properties if
the removal of the surface states (i.e.~the passivation) indeed fails.

We note that in comparison to the bulk calculation in Ref.~\cite{koenig2021a}, the effect of many-body
interaction corrections resembles much more that of a scissors operator (see
Supplemental Material \cite{supplemental}).


\section{Electronic transport properties of the films}
We used the same interpolation method as described above to obtain the density
of states and conductivity $\sigma$ of the slabs.
Since the effect of confinement on the surface states apparently is described reasonably
well with density functional theory (at least at $\overline{\text{M}}$ where we compared with experimental data),
the DFT states were used as basis for the following calculations.
Our code uses the tetrahedron method for the calculation of the DOS
and the Boltzmann transport equation with the constant relaxation time
approximation for $\sigma$, similar to \textsc{BoltzTraP2} \cite{madsen2006,madsen2018}.
The conductivity is then linearly proportional to the constant relaxation time parameter $\tau$.
When calculating $\sigma$ for different temperatures, the occupation of
the states changes. Changes of the scattering rates that have been used
to distinguish between the metallic surface and the semiconducting bulk
channel of the films in experimental data (see, e.g., Ref.~\cite{fei2010})
can however not be considered in this simple model.

Figure \ref{fig:dosAndSigma} (a) shows the density of states of the
thin films.
A good description of the surface states is particularly important
since they constitute the majority of states at the Fermi level.
As we already know, only the three bilayer thick film
has a band gap. Notably, films with 10, 15, and 30 bilayers
all have approximately the same feature at the Fermi level,
which originates from the surface states and thus does not appreciably change
if the film thickness is increased further.
Upon normalization with the number of bilayers
in Fig.~\ref{fig:dosAndSigma} (c),
the density of valence states of 10, 15,
and 30 bilayer thick films approaches a
converged value, i.e.~the bulk limit.
A prominent feature at around \SIrange{100}{200}{\milli\electronvolt}
above the Fermi level can be observed for the three bilayer thick film
and corresponds to an almost flat surface band.
The curvature of this band increases in thicker films but
otherwise does not change its position in energy so that
the peak spreads over a wider energy range.

The observations above are useful for the interpretation of the
conductivity, which is shown in Figs.~\ref{fig:dosAndSigma} (b)
and \ref{fig:dosAndSigma} (d) without and with normalization with the slab thickness.
First of all, the three and five bilayer thick films
are insulating at \SI{0}{\kelvin}. The low-temperature behavior of the sheet conductivity,
Fig.~\ref{fig:dosAndSigma} (b), clearly shows that
the critical thickness, below which the surface states
require thermal excitation in order to conduct current,
lies between five and ten bilayers.
In contrast to that, the thicker films all have finite conductivity
for low temperatures since the surface states are already partially filled.
The $10$ and $15$ bilayer films have a very similar conductivity
at \SI{0}{\kelvin} and the core of the film remains insulating.
The thickest slab with $30$ bilayers has an additional low-temperature contribution
from the $\overline{\Gamma}$ valence bands.

Similar to Fig.~4 in Ref.~\cite{xiao2012}, when normalized with the
number of bilayers, thicker films have a lower conductivity at low
temperatures before thermal excitation makes the semiconducting core conduct.
Before that, only the surface contributes to the current.
However, the films in Fig.~\ref{fig:dosAndSigma} (d) are thinner than those used in the experiment
and we observe the transition from the semiconducting
to semimetallic films already at around five bilayers.
The G$_0$W$_0$ corrected band structures in
Fig.~\ref{fig:G0W0Bands} suggest the same since the energy
gap here also closes at around five bilayers. This threshold agrees well
with the measurements in Ref.~\cite{hirahara2007}.
Aitani \textit{et al.}~\cite{aitani2014} report that a transition between
surface and bulk dominated transport happens above a film thickness
of 16 bilayers.

We expect that more sophisticated models including temperature-dependent
scattering will further improve the agreement with
the measurements. The changes in the scattering are important
to show where the surface contribution is outweighed by the
semiconducting core because the metallic states at the surface
conduct less at higher temperatures, so that it is possible to
clearly distinguish the two channels by fitting $\sigma(T)$
where required, see, e.g.,Ref.~\cite{xiao2012}.
With the simpler approach presented here, we have already
identified the onset of conductivity in the unterminated films
as a function of thickness
in order to compare to the experimentally predicted
semimetal-to-semiconductor transition.
We also note that for these very thin films an allotropic phase
exists, which has a similar effect on the conductivity, see,
e.g., Refs.~\cite{nagao2004,hirahara2007,xiao2012} for more details.


\section{Discussion}
In the first part of this paper we have shown how the surface states
on thin Bi\hkl(1 1 1) slabs react to different kinds of surface termination.
The metallic states are not removed but shifted at the three main symmetry
points $\overline{\text{K}}$, $\overline{\Gamma}$, and $\overline{\text{M}}$. The importance of the $\overline{\Gamma}$ point for a global
band gap as well as relaxation effects (inter-bilayer spacing and reorientations
due to surface termination) have been highlighted. We found that the layered-oxide
model shows negligible interaction with the film and thus does not
passivate the surface. In summary, the passivation of the surface of a three
bilayer film was not possible in DFT without additional constraints
on the structure.

Subsequently, we have compared the performance of DFT and G$_0$W$_0$ regarding
an accurate description of the confinement effects in thin films. The DFT
calculations agree very well with experimental results in the literature
although the slab thickness is not directly comparable. G$_0$W$_0$, which
also reproduces the direct proportionality of the $\overline{\text{M}}$ gap with the reciprocal
thickness, overestimates the splitting of the bands. This may be a result
of substrate-induced strain, which we did not account for above.
In a test calculation for the film in Fig.~\ref{fig:relaxationAndStrainEffects}
(b) we found the $\overline{\text{M}}$ gap to increase slightly from \SIrange{1.25}{1.28}{\electronvolt}
upon applying the \SI{1.3}{\percent} compressive
strain. Thus, it is likely that the difference in the $\overline{\text{M}}$ gap is rather due
to the vicinity of the substrate and the corresponding effect on the screening
in the thin film in experiment.
Substantial changes in the screening due to a substrate and a concomitant
decrease of the band gap has been shown for thin films in Ref.~\cite{qiu2017}.
The good agreement between the experimental $\overline{\text{M}}$ gap and the density functional
theory results justifies the use of DFT for this passivation study.

Since the bulk material is topologically trivial in DFT, those films, which
have the same atomic structure do not have to have topological surface states,
which can not be passivated.
The situation is different once the many-body corrections are considered.
Using the same methods, we have shown previously \cite{koenig2021a} that bismuth
is topologically nontrivial. If bismuth was an insulator, this would mean
that metallic surface states exist on the Bi\hkl(1 1 1) surface, which
connect the valence and conduction bands. Nevertheless, the band structure
of the three bilayer thick film in Fig.~\ref{fig:G0W0Bands} has a global band
gap. There may be two reasons for this kind of behavior. Firstly, as was
pointed out in Ref.~\cite{ohtsubo2013},
the overlapping bulk valence and conduction
bands can be connected by surface states without crossing the Fermi level.
Explicitly, this would mean that both surface states connect to the bulk valence
bands at $\overline{\Gamma}$ with a local energy gap between them and further connect
to the valence and conduction bands at $\overline{\text{M}}$.
Secondly, confinement is a further perturbation to the system. In particular,
although it is possible to construct the bulk system corresponding to a strained
film, it is not as straightforward to construct such a system taking into
account the effects of confinement. Strain is known to drive topological
phase transitions and we assume that this is the case for confinement as well,
so that the topological constraints do not necessarily remain in very thin films.

With a simple model we have investigated the conductivity of the films
and compared $\sigma$ of the calculated band structures to the popular empirical
two-channel model. We observe an onset of the surface state conductivity
for a thickness above five bilayers, which agrees well with experimental results.
For the interpretation of experimental
data for films with a thickness of only a few bilayers, the semiconducting
allotropic phase also has to be considered, see, e.g., Refs.~\cite{hirahara2007,xiao2012}.
A clear distinction between the metallic surface states and the semiconducting
core or bulk states, which are populated by thermal excitation, requires
the inclusion of more details like electron-phonon scattering.

Realistically, the relaxation time $\tau$ depends on the temperature
as well as the band energy and position in reciprocal space. Thus,
it is also a function of the thickness of the film.
Further relaxations of the thin films that may alter the electronic
structure have been neglected completely.
Disregarding all these complexities, we can estimate $\tau$ for the
surface states by comparing with the experimentally obtained
surface state conductivity of approximately \SI{1.5e-3}{\per\ohm\sq} \cite{hirahara2007}.
The calculated low temperature values for the sheet conductivity
between $100$ to $\SI{250}{\per\ohm\sq}/\tau[\si{\nano\second}]$
correspond to a relaxation time of only \SIrange{6}{15}{\femto\second}
which appears to be short and is much smaller than the estimated value
of \SI{200}{\femto\second} which was reported in Ref.~\cite{du2016}.
However, the relaxation time is comparable to the bulk value reported
in Ref.~\cite{omahony2019}. There the electronic lifetime was found to be
about the lifetime of the electronic force driving the E$_g$ vibrational mode
which was independent of the excitation energy (corresponding to the band gap
for the electron-hole recombination, changing with confinement).

There have been few published studies concerning the effect of termination
on the Bi\hkl(1 1 1) surface.
Liu \textit{et al.}~have schematically shown in Ref.~\cite{liu2011}
how a hydrogen termination changes the band structure.
Based on the similarity of their results to Fig.~\ref{fig:simpleTerminationHandOH} (a),
we assume that they have also only relaxed the H atoms on the surface
but did not allow the whole film to relax. The shown band structure has no substantial
band gap.
The authors of Ref.~\cite{chang2019} investigated the effect of a
surface potential, mainly in the context of band topology and the
system's response to perturbations.
An asymmetric potential was used there to remove some of the interaction
between the surface states at $\overline{\text{M}}$ and to thus distinguish the topologically
trivial and nontrivial phases. The surface potential was then replaced with surface hydrogen bonding
to a similar effect as in our calculations and the results of Ref.~\cite{liu2011}.
Experimentally, arguments for a passivation by oxidation have been found in Ref.~\cite{hirahara2007},
although temperature-dependent measurements would be particularly useful in this context,
e.g., in order to exclude the full oxidation of the six bilayer thick film.
The temperature dependence in the two-channel model and the corresponding measurements
only consider unterminated surfaces.
The atomistic modeling of the oxide and its interaction with the film as presented above
does not suggest a passivation effect.
In summary, with the results presented in this paper, we have extended
the knowledge on the passivation of thin Bi\hkl(1 1 1) films with different kinds of termination
and have pointed out the side effects, which occur due to changes
in the crystal structure.


\begin{acknowledgments}
\section{Acknowledgments}
This work has been funded by Science Foundation Ireland through the Principal Investigator Award No.~13/IA/1956.
The authors wish to acknowledge the Irish Centre for High-End Computing (ICHEC) for the provision of computational facilities and support.
Support is also provided by the Nottingham Ningbo New Materials Institute and the National Natural Science Foundation of China with Project Code 61974079.
Atomistic structures were visualized with the \textsc{Vesta} software \cite{momma2011}.
\end{acknowledgments}

\bibliography{./bibliography}

\begin{thebibliography}{79}%
\makeatletter
\providecommand \@ifxundefined [1]{%
 \@ifx{#1\undefined}
}%
\providecommand \@ifnum [1]{%
 \ifnum #1\expandafter \@firstoftwo
 \else \expandafter \@secondoftwo
 \fi
}%
\providecommand \@ifx [1]{%
 \ifx #1\expandafter \@firstoftwo
 \else \expandafter \@secondoftwo
 \fi
}%
\providecommand \natexlab [1]{#1}%
\providecommand \enquote  [1]{``#1''}%
\providecommand \bibnamefont  [1]{#1}%
\providecommand \bibfnamefont [1]{#1}%
\providecommand \citenamefont [1]{#1}%
\providecommand \href@noop [0]{\@secondoftwo}%
\providecommand \href [0]{\begingroup \@sanitize@url \@href}%
\providecommand \@href[1]{\@@startlink{#1}\@@href}%
\providecommand \@@href[1]{\endgroup#1\@@endlink}%
\providecommand \@sanitize@url [0]{\catcode `\\12\catcode `\$12\catcode
  `\&12\catcode `\#12\catcode `\^12\catcode `\_12\catcode `\%12\relax}%
\providecommand \@@startlink[1]{}%
\providecommand \@@endlink[0]{}%
\providecommand \url  [0]{\begingroup\@sanitize@url \@url }%
\providecommand \@url [1]{\endgroup\@href {#1}{\urlprefix }}%
\providecommand \urlprefix  [0]{URL }%
\providecommand \Eprint [0]{\href }%
\providecommand \doibase [0]{http://dx.doi.org/}%
\providecommand \selectlanguage [0]{\@gobble}%
\providecommand \bibinfo  [0]{\@secondoftwo}%
\providecommand \bibfield  [0]{\@secondoftwo}%
\providecommand \translation [1]{[#1]}%
\providecommand \BibitemOpen [0]{}%
\providecommand \bibitemStop [0]{}%
\providecommand \bibitemNoStop [0]{.\EOS\space}%
\providecommand \EOS [0]{\spacefactor3000\relax}%
\providecommand \BibitemShut  [1]{\csname bibitem#1\endcsname}%
\let\auto@bib@innerbib\@empty
\bibitem [{\citenamefont {Aguilera}\ \emph {et~al.}(2015)\citenamefont
  {Aguilera}, \citenamefont {Friedrich},\ and\ \citenamefont
  {Blügel}}]{aguilera2015}%
  \BibitemOpen
  \bibfield  {author} {\bibinfo {author} {\bibfnamefont {I.}~\bibnamefont
  {Aguilera}}, \bibinfo {author} {\bibfnamefont {C.}~\bibnamefont {Friedrich}},
  \ and\ \bibinfo {author} {\bibfnamefont {S.}~\bibnamefont {Blügel}},\
  }\bibfield  {title} {\enquote {\bibinfo {title} {Electronic phase transitions
  of bismuth under strain from relativistic self-consistent {GW}
  calculations},}\ }\href@noop {} {\bibfield  {journal} {\bibinfo  {journal}
  {Phys. Rev. B}\ }\textbf {\bibinfo {volume} {91}},\ \bibinfo {pages} {125129}
  (\bibinfo {year} {2015})}\BibitemShut {NoStop}%
\bibitem [{\citenamefont {Édel'man}(1975)}]{edelman1975}%
  \BibitemOpen
  \bibfield  {author} {\bibinfo {author} {\bibfnamefont {V.~S.}\ \bibnamefont
  {Édel'man}},\ }\bibfield  {title} {\enquote {\bibinfo {title} {Investigation
  of bismuth in a quantizing field},}\ }\href@noop {} {\bibfield  {journal}
  {\bibinfo  {journal} {J. Exp. Theor. Phys.}\ }\textbf {\bibinfo {volume}
  {41}},\ \bibinfo {pages} {125} (\bibinfo {year} {1975})}\BibitemShut
  {NoStop}%
\bibitem [{\citenamefont {Isaacson}\ and\ \citenamefont
  {Williams}(1969)}]{isaacson1969b}%
  \BibitemOpen
  \bibfield  {author} {\bibinfo {author} {\bibfnamefont {R.~T.}\ \bibnamefont
  {Isaacson}}\ and\ \bibinfo {author} {\bibfnamefont {G.~A.}\ \bibnamefont
  {Williams}},\ }\bibfield  {title} {\enquote {\bibinfo {title} {Alfvén-wave
  propagation in solid-state plasmas. {III}. {Quantum} oscillations of the
  {Fermi} surface of bismuth},}\ }\href@noop {} {\bibfield  {journal} {\bibinfo
   {journal} {Phys. Rev.}\ }\textbf {\bibinfo {volume} {185}},\ \bibinfo
  {pages} {682} (\bibinfo {year} {1969})}\BibitemShut {NoStop}%
\bibitem [{\citenamefont {Dinger}\ and\ \citenamefont
  {Lawson}(1973)}]{dinger1973}%
  \BibitemOpen
  \bibfield  {author} {\bibinfo {author} {\bibfnamefont {R.~J.}\ \bibnamefont
  {Dinger}}\ and\ \bibinfo {author} {\bibfnamefont {A.~W.}\ \bibnamefont
  {Lawson}},\ }\bibfield  {title} {\enquote {\bibinfo {title} {Cyclotron
  resonance and the {Cohen} nonellipsoidal nonparabolic model for bismuth.
  {III}. {Experimental} results},}\ }\href@noop {} {\bibfield  {journal}
  {\bibinfo  {journal} {Phys. Rev. B}\ }\textbf {\bibinfo {volume} {7}},\
  \bibinfo {pages} {5215} (\bibinfo {year} {1973})}\BibitemShut {NoStop}%
\bibitem [{\citenamefont {Smith}\ \emph {et~al.}(1964)\citenamefont {Smith},
  \citenamefont {Baraff},\ and\ \citenamefont {Rowell}}]{smith1964}%
  \BibitemOpen
  \bibfield  {author} {\bibinfo {author} {\bibfnamefont {G.~E.}\ \bibnamefont
  {Smith}}, \bibinfo {author} {\bibfnamefont {G.~A.}\ \bibnamefont {Baraff}}, \
  and\ \bibinfo {author} {\bibfnamefont {J.~M.}\ \bibnamefont {Rowell}},\
  }\bibfield  {title} {\enquote {\bibinfo {title} {Effective $g$ factor of
  electrons and holes in bismuth},}\ }\href@noop {} {\bibfield  {journal}
  {\bibinfo  {journal} {Phys. Rev.}\ }\textbf {\bibinfo {volume} {135}},\
  \bibinfo {pages} {A1118} (\bibinfo {year} {1964})}\BibitemShut {NoStop}%
\bibitem [{\citenamefont {Maltz}\ and\ \citenamefont
  {Dresselhaus}(1970)}]{maltz1970}%
  \BibitemOpen
  \bibfield  {author} {\bibinfo {author} {\bibfnamefont {M.}~\bibnamefont
  {Maltz}}\ and\ \bibinfo {author} {\bibfnamefont {M.~S.}\ \bibnamefont
  {Dresselhaus}},\ }\bibfield  {title} {\enquote {\bibinfo {title}
  {Magnetoreflection studies in bismuth},}\ }\href@noop {} {\bibfield
  {journal} {\bibinfo  {journal} {Phys. Rev. B}\ }\textbf {\bibinfo {volume}
  {2}},\ \bibinfo {pages} {2877} (\bibinfo {year} {1970})}\BibitemShut
  {NoStop}%
\bibitem [{\citenamefont {Vecchi}\ and\ \citenamefont
  {Dresselhaus}(1974)}]{vecchi1974}%
  \BibitemOpen
  \bibfield  {author} {\bibinfo {author} {\bibfnamefont {M.~P.}\ \bibnamefont
  {Vecchi}}\ and\ \bibinfo {author} {\bibfnamefont {M.~S.}\ \bibnamefont
  {Dresselhaus}},\ }\bibfield  {title} {\enquote {\bibinfo {title} {Temperature
  dependence of the band parameters of bismuth},}\ }\href@noop {} {\bibfield
  {journal} {\bibinfo  {journal} {Phys. Rev. B}\ }\textbf {\bibinfo {volume}
  {10}},\ \bibinfo {pages} {771} (\bibinfo {year} {1974})}\BibitemShut
  {NoStop}%
\bibitem [{\citenamefont {Brown}\ \emph {et~al.}(1963)\citenamefont {Brown},
  \citenamefont {Mavroides},\ and\ \citenamefont {Lax}}]{brown1963}%
  \BibitemOpen
  \bibfield  {author} {\bibinfo {author} {\bibfnamefont {R.~N.}\ \bibnamefont
  {Brown}}, \bibinfo {author} {\bibfnamefont {J.~G.}\ \bibnamefont
  {Mavroides}}, \ and\ \bibinfo {author} {\bibfnamefont {B.}~\bibnamefont
  {Lax}},\ }\bibfield  {title} {\enquote {\bibinfo {title} {Magnetoreflection
  in bismuth},}\ }\href@noop {} {\bibfield  {journal} {\bibinfo  {journal}
  {Phys. Rev.}\ }\textbf {\bibinfo {volume} {129}},\ \bibinfo {pages} {2055}
  (\bibinfo {year} {1963})}\BibitemShut {NoStop}%
\bibitem [{\citenamefont {Hofmann}(2006)}]{hofmann2006}%
  \BibitemOpen
  \bibfield  {author} {\bibinfo {author} {\bibfnamefont {P.}~\bibnamefont
  {Hofmann}},\ }\bibfield  {title} {\enquote {\bibinfo {title} {The surfaces of
  bismuth: Structural and electronic properties},}\ }\href@noop {} {\bibfield
  {journal} {\bibinfo  {journal} {Prog. Surf. Sci.}\ }\textbf {\bibinfo
  {volume} {81}},\ \bibinfo {pages} {191} (\bibinfo {year} {2006})}\BibitemShut
  {NoStop}%
\bibitem [{\citenamefont {Hirahara}\ \emph {et~al.}(2006)\citenamefont
  {Hirahara}, \citenamefont {Nagao}, \citenamefont {Matsuda}, \citenamefont
  {Bihlmayer}, \citenamefont {Chulkov}, \citenamefont {Koroteev}, \citenamefont
  {Echenique}, \citenamefont {Saito},\ and\ \citenamefont
  {Hasegawa}}]{hirahara2006}%
  \BibitemOpen
  \bibfield  {author} {\bibinfo {author} {\bibfnamefont {T.}~\bibnamefont
  {Hirahara}}, \bibinfo {author} {\bibfnamefont {T.}~\bibnamefont {Nagao}},
  \bibinfo {author} {\bibfnamefont {I.}~\bibnamefont {Matsuda}}, \bibinfo
  {author} {\bibfnamefont {G.}~\bibnamefont {Bihlmayer}}, \bibinfo {author}
  {\bibfnamefont {E.~V.}\ \bibnamefont {Chulkov}}, \bibinfo {author}
  {\bibfnamefont {Y.~M.}\ \bibnamefont {Koroteev}}, \bibinfo {author}
  {\bibfnamefont {P.~M.}\ \bibnamefont {Echenique}}, \bibinfo {author}
  {\bibfnamefont {M.}~\bibnamefont {Saito}}, \ and\ \bibinfo {author}
  {\bibfnamefont {S.}~\bibnamefont {Hasegawa}},\ }\bibfield  {title} {\enquote
  {\bibinfo {title} {Role of spin-orbit coupling and hybridization effects in
  the electronic structure of ultrathin {Bi} films},}\ }\href@noop {}
  {\bibfield  {journal} {\bibinfo  {journal} {Phys. Rev. Lett.}\ }\textbf
  {\bibinfo {volume} {97}},\ \bibinfo {pages} {146803} (\bibinfo {year}
  {2006})}\BibitemShut {NoStop}%
\bibitem [{\citenamefont {Hirahara}\ \emph
  {et~al.}(2007{\natexlab{a}})\citenamefont {Hirahara}, \citenamefont {Nagao},
  \citenamefont {Matsuda}, \citenamefont {Bihlmayer}, \citenamefont {Chulkov},
  \citenamefont {Koroteev},\ and\ \citenamefont {Hasegawa}}]{hirahara2007b}%
  \BibitemOpen
  \bibfield  {author} {\bibinfo {author} {\bibfnamefont {T.}~\bibnamefont
  {Hirahara}}, \bibinfo {author} {\bibfnamefont {T.}~\bibnamefont {Nagao}},
  \bibinfo {author} {\bibfnamefont {I.}~\bibnamefont {Matsuda}}, \bibinfo
  {author} {\bibfnamefont {G.}~\bibnamefont {Bihlmayer}}, \bibinfo {author}
  {\bibfnamefont {E.~V.}\ \bibnamefont {Chulkov}}, \bibinfo {author}
  {\bibfnamefont {Y.~M.}\ \bibnamefont {Koroteev}}, \ and\ \bibinfo {author}
  {\bibfnamefont {S.}~\bibnamefont {Hasegawa}},\ }\bibfield  {title} {\enquote
  {\bibinfo {title} {Quantum well states in ultrathin {Bi} films:
  {Angle}-resolved photoemission spectroscopy and first-principles calculations
  study},}\ }\href@noop {} {\bibfield  {journal} {\bibinfo  {journal} {Phys.
  Rev. B}\ }\textbf {\bibinfo {volume} {75}},\ \bibinfo {pages} {035422}
  (\bibinfo {year} {2007}{\natexlab{a}})}\BibitemShut {NoStop}%
\bibitem [{\citenamefont {Ishida}(2017)}]{ishida2016}%
  \BibitemOpen
  \bibfield  {author} {\bibinfo {author} {\bibfnamefont {H.}~\bibnamefont
  {Ishida}},\ }\bibfield  {title} {\enquote {\bibinfo {title} {Decay length of
  surface-state wave functions on {Bi}\hkl(1 1 1)},}\ }\href@noop {} {\bibfield
   {journal} {\bibinfo  {journal} {J. Phys.: Condens. Matter}\ }\textbf
  {\bibinfo {volume} {29}},\ \bibinfo {pages} {015002} (\bibinfo {year}
  {2017})}\BibitemShut {NoStop}%
\bibitem [{\citenamefont {Ito}\ \emph {et~al.}(2016)\citenamefont {Ito},
  \citenamefont {Feng}, \citenamefont {Arita}, \citenamefont {Takayama},
  \citenamefont {Liu}, \citenamefont {Someya}, \citenamefont {Chen},
  \citenamefont {Iimori}, \citenamefont {Namatame}, \citenamefont {Taniguchi},
  \citenamefont {Cheng}, \citenamefont {Tang}, \citenamefont {Komori},
  \citenamefont {Kobayashi}, \citenamefont {Chiang},\ and\ \citenamefont
  {Matsuda}}]{ito2016}%
  \BibitemOpen
  \bibfield  {author} {\bibinfo {author} {\bibfnamefont {S.}~\bibnamefont
  {Ito}}, \bibinfo {author} {\bibfnamefont {B.}~\bibnamefont {Feng}}, \bibinfo
  {author} {\bibfnamefont {M.}~\bibnamefont {Arita}}, \bibinfo {author}
  {\bibfnamefont {A.}~\bibnamefont {Takayama}}, \bibinfo {author}
  {\bibfnamefont {R.-Y.}\ \bibnamefont {Liu}}, \bibinfo {author} {\bibfnamefont
  {T.}~\bibnamefont {Someya}}, \bibinfo {author} {\bibfnamefont {W.-C.}\
  \bibnamefont {Chen}}, \bibinfo {author} {\bibfnamefont {T.}~\bibnamefont
  {Iimori}}, \bibinfo {author} {\bibfnamefont {H.}~\bibnamefont {Namatame}},
  \bibinfo {author} {\bibfnamefont {M.}~\bibnamefont {Taniguchi}}, \bibinfo
  {author} {\bibfnamefont {C.-M.}\ \bibnamefont {Cheng}}, \bibinfo {author}
  {\bibfnamefont {S.-J.}\ \bibnamefont {Tang}}, \bibinfo {author}
  {\bibfnamefont {F.}~\bibnamefont {Komori}}, \bibinfo {author} {\bibfnamefont
  {K.}~\bibnamefont {Kobayashi}}, \bibinfo {author} {\bibfnamefont {T.-C.}\
  \bibnamefont {Chiang}}, \ and\ \bibinfo {author} {\bibfnamefont
  {I.}~\bibnamefont {Matsuda}},\ }\bibfield  {title} {\enquote {\bibinfo
  {title} {Proving nontrivial topology of pure bismuth by quantum
  confinement},}\ }\href@noop {} {\bibfield  {journal} {\bibinfo  {journal}
  {Phys. Rev. Lett.}\ }\textbf {\bibinfo {volume} {117}},\ \bibinfo {pages}
  {236402} (\bibinfo {year} {2016})}\BibitemShut {NoStop}%
\bibitem [{\citenamefont {Cantele}\ and\ \citenamefont
  {Ninno}(2017)}]{cantele2017}%
  \BibitemOpen
  \bibfield  {author} {\bibinfo {author} {\bibfnamefont {G.}~\bibnamefont
  {Cantele}}\ and\ \bibinfo {author} {\bibfnamefont {D.}~\bibnamefont
  {Ninno}},\ }\bibfield  {title} {\enquote {\bibinfo {title} {Size-dependent
  structural and electronic properties of {Bi}\hkl(1 1 1) ultrathin nanofilms
  from first principles},}\ }\href@noop {} {\bibfield  {journal} {\bibinfo
  {journal} {Phys. Rev. Materials}\ }\textbf {\bibinfo {volume} {1}},\ \bibinfo
  {pages} {014002} (\bibinfo {year} {2017})}\BibitemShut {NoStop}%
\bibitem [{\citenamefont {Lutskii}(1965)}]{lutskii1965}%
  \BibitemOpen
  \bibfield  {author} {\bibinfo {author} {\bibfnamefont {V.~N.}\ \bibnamefont
  {Lutskii}},\ }\bibfield  {title} {\enquote {\bibinfo {title} {Features of
  optical absorption of metallic films in the region where the metal turns into
  a dielectric},}\ }\href@noop {} {\bibfield  {journal} {\bibinfo  {journal}
  {J. Exp. Theor. Phys. Lett.}\ }\textbf {\bibinfo {volume} {2}},\ \bibinfo
  {pages} {245} (\bibinfo {year} {1965})}\BibitemShut {NoStop}%
\bibitem [{\citenamefont {Ogrin}\ \emph {et~al.}(1966)\citenamefont {Ogrin},
  \citenamefont {Lutskii},\ and\ \citenamefont {Elinson}}]{ogrin1966}%
  \BibitemOpen
  \bibfield  {author} {\bibinfo {author} {\bibfnamefont {Y.~F.}\ \bibnamefont
  {Ogrin}}, \bibinfo {author} {\bibfnamefont {V.~N.}\ \bibnamefont {Lutskii}},
  \ and\ \bibinfo {author} {\bibfnamefont {M.~I.}\ \bibnamefont {Elinson}},\
  }\bibfield  {title} {\enquote {\bibinfo {title} {Observation of quantum size
  effects in thin bismuth films},}\ }\href@noop {} {\bibfield  {journal}
  {\bibinfo  {journal} {J. Exp. Theor. Phys. Lett.}\ }\textbf {\bibinfo
  {volume} {3}},\ \bibinfo {pages} {71} (\bibinfo {year} {1966})}\BibitemShut
  {NoStop}%
\bibitem [{\citenamefont {Sandomirskii}(1967)}]{sandomirskii1967}%
  \BibitemOpen
  \bibfield  {author} {\bibinfo {author} {\bibfnamefont {V.~B.}\ \bibnamefont
  {Sandomirskii}},\ }\bibfield  {title} {\enquote {\bibinfo {title} {Quantum
  size effect in a semimetal film},}\ }\href@noop {} {\bibfield  {journal}
  {\bibinfo  {journal} {J. Exp. Theor. Phys.}\ }\textbf {\bibinfo {volume}
  {25}},\ \bibinfo {pages} {101} (\bibinfo {year} {1967})}\BibitemShut
  {NoStop}%
\bibitem [{\citenamefont {Hoffman}\ \emph {et~al.}(1993)\citenamefont
  {Hoffman}, \citenamefont {Meyer}, \citenamefont {Bartoli}, \citenamefont
  {Di~Venere}, \citenamefont {Yi}, \citenamefont {Hou}, \citenamefont {Wang},
  \citenamefont {Ketterson},\ and\ \citenamefont {Wong}}]{hoffman1993}%
  \BibitemOpen
  \bibfield  {author} {\bibinfo {author} {\bibfnamefont {C.~A.}\ \bibnamefont
  {Hoffman}}, \bibinfo {author} {\bibfnamefont {J.~R.}\ \bibnamefont {Meyer}},
  \bibinfo {author} {\bibfnamefont {F.~J.}\ \bibnamefont {Bartoli}}, \bibinfo
  {author} {\bibfnamefont {A.}~\bibnamefont {Di~Venere}}, \bibinfo {author}
  {\bibfnamefont {X.~J.}\ \bibnamefont {Yi}}, \bibinfo {author} {\bibfnamefont
  {C.~L.}\ \bibnamefont {Hou}}, \bibinfo {author} {\bibfnamefont {H.~C.}\
  \bibnamefont {Wang}}, \bibinfo {author} {\bibfnamefont {J.~B.}\ \bibnamefont
  {Ketterson}}, \ and\ \bibinfo {author} {\bibfnamefont {G.~K.}\ \bibnamefont
  {Wong}},\ }\bibfield  {title} {\enquote {\bibinfo {title}
  {Semimetal-to-semiconductor transition in bismuth thin films},}\ }\href@noop
  {} {\bibfield  {journal} {\bibinfo  {journal} {Phys. Rev. B}\ }\textbf
  {\bibinfo {volume} {48}},\ \bibinfo {pages} {11431} (\bibinfo {year}
  {1993})}\BibitemShut {NoStop}%
\bibitem [{\citenamefont {Rogacheva}\ \emph {et~al.}(2008)\citenamefont
  {Rogacheva}, \citenamefont {Lyubchenko},\ and\ \citenamefont
  {Dresselhaus}}]{rogacheva2008}%
  \BibitemOpen
  \bibfield  {author} {\bibinfo {author} {\bibfnamefont {E.~I.}\ \bibnamefont
  {Rogacheva}}, \bibinfo {author} {\bibfnamefont {S.~G.}\ \bibnamefont
  {Lyubchenko}}, \ and\ \bibinfo {author} {\bibfnamefont {M.~S.}\ \bibnamefont
  {Dresselhaus}},\ }\bibfield  {title} {\enquote {\bibinfo {title}
  {Semimetal-semiconductor transition in thin {Bi} films},}\ }\href@noop {}
  {\bibfield  {journal} {\bibinfo  {journal} {Thin Solid Films}\ }\textbf
  {\bibinfo {volume} {516}},\ \bibinfo {pages} {3411} (\bibinfo {year}
  {2008})}\BibitemShut {NoStop}%
\bibitem [{\citenamefont {Gity}\ \emph {et~al.}(2017)\citenamefont {Gity},
  \citenamefont {Ansari}, \citenamefont {Lanius}, \citenamefont {Schüffelgen},
  \citenamefont {Mussler}, \citenamefont {Grützmacher},\ and\ \citenamefont
  {Greer}}]{gity2017}%
  \BibitemOpen
  \bibfield  {author} {\bibinfo {author} {\bibfnamefont {F.}~\bibnamefont
  {Gity}}, \bibinfo {author} {\bibfnamefont {L.}~\bibnamefont {Ansari}},
  \bibinfo {author} {\bibfnamefont {M.}~\bibnamefont {Lanius}}, \bibinfo
  {author} {\bibfnamefont {P.}~\bibnamefont {Schüffelgen}}, \bibinfo {author}
  {\bibfnamefont {G.}~\bibnamefont {Mussler}}, \bibinfo {author} {\bibfnamefont
  {D.}~\bibnamefont {Grützmacher}}, \ and\ \bibinfo {author} {\bibfnamefont
  {J.~C.}\ \bibnamefont {Greer}},\ }\bibfield  {title} {\enquote {\bibinfo
  {title} {Reinventing solid state electronics: Harnessing quantum confinement
  in bismuth thin films},}\ }\href@noop {} {\bibfield  {journal} {\bibinfo
  {journal} {Appl. Phys. Lett.}\ }\textbf {\bibinfo {volume} {110}},\ \bibinfo
  {pages} {093111} (\bibinfo {year} {2017})}\BibitemShut {NoStop}%
\bibitem [{\citenamefont {Gity}\ \emph {et~al.}(2018)\citenamefont {Gity},
  \citenamefont {Ansari}, \citenamefont {König}, \citenamefont {Verni},
  \citenamefont {Holmes}, \citenamefont {Long}, \citenamefont {Lanius},
  \citenamefont {Schüffelgen}, \citenamefont {Mussler}, \citenamefont
  {Grützmacher},\ and\ \citenamefont {Greer}}]{gity2018}%
  \BibitemOpen
  \bibfield  {author} {\bibinfo {author} {\bibfnamefont {F.}~\bibnamefont
  {Gity}}, \bibinfo {author} {\bibfnamefont {L.}~\bibnamefont {Ansari}},
  \bibinfo {author} {\bibfnamefont {C.}~\bibnamefont {König}}, \bibinfo
  {author} {\bibfnamefont {G.~A.}\ \bibnamefont {Verni}}, \bibinfo {author}
  {\bibfnamefont {J.~D.}\ \bibnamefont {Holmes}}, \bibinfo {author}
  {\bibfnamefont {B.}~\bibnamefont {Long}}, \bibinfo {author} {\bibfnamefont
  {M.}~\bibnamefont {Lanius}}, \bibinfo {author} {\bibfnamefont
  {P.}~\bibnamefont {Schüffelgen}}, \bibinfo {author} {\bibfnamefont
  {G.}~\bibnamefont {Mussler}}, \bibinfo {author} {\bibfnamefont
  {D.}~\bibnamefont {Grützmacher}}, \ and\ \bibinfo {author} {\bibfnamefont
  {J.~C.}\ \bibnamefont {Greer}},\ }\bibfield  {title} {\enquote {\bibinfo
  {title} {Metal-semimetal {Schottky} diode relying on quantum confinement},}\
  }\href@noop {} {\bibfield  {journal} {\bibinfo  {journal} {Microelectron.
  Eng.}\ }\textbf {\bibinfo {volume} {195}},\ \bibinfo {pages} {21} (\bibinfo
  {year} {2018})}\BibitemShut {NoStop}%
\bibitem [{\citenamefont {Hirahara}\ \emph
  {et~al.}(2007{\natexlab{b}})\citenamefont {Hirahara}, \citenamefont
  {Matsuda}, \citenamefont {Yamazaki}, \citenamefont {Miyata}, \citenamefont
  {Hasegawa},\ and\ \citenamefont {Nagao}}]{hirahara2007}%
  \BibitemOpen
  \bibfield  {author} {\bibinfo {author} {\bibfnamefont {T.}~\bibnamefont
  {Hirahara}}, \bibinfo {author} {\bibfnamefont {I.}~\bibnamefont {Matsuda}},
  \bibinfo {author} {\bibfnamefont {S.}~\bibnamefont {Yamazaki}}, \bibinfo
  {author} {\bibfnamefont {N.}~\bibnamefont {Miyata}}, \bibinfo {author}
  {\bibfnamefont {S.}~\bibnamefont {Hasegawa}}, \ and\ \bibinfo {author}
  {\bibfnamefont {T.}~\bibnamefont {Nagao}},\ }\bibfield  {title} {\enquote
  {\bibinfo {title} {Large surface-state conductivity in ultrathin {Bi}
  films},}\ }\href@noop {} {\bibfield  {journal} {\bibinfo  {journal} {Appl.
  Phys. Lett.}\ }\textbf {\bibinfo {volume} {91}},\ \bibinfo {pages} {202106}
  (\bibinfo {year} {2007}{\natexlab{b}})}\BibitemShut {NoStop}%
\bibitem [{\citenamefont {Pang}\ \emph {et~al.}(2010)\citenamefont {Pang},
  \citenamefont {Liang}, \citenamefont {Liao}, \citenamefont {Yin},\ and\
  \citenamefont {Chen}}]{fei2010}%
  \BibitemOpen
  \bibfield  {author} {\bibinfo {author} {\bibfnamefont {F.}~\bibnamefont
  {Pang}}, \bibinfo {author} {\bibfnamefont {X.-J.}\ \bibnamefont {Liang}},
  \bibinfo {author} {\bibfnamefont {Z.-L.}\ \bibnamefont {Liao}}, \bibinfo
  {author} {\bibfnamefont {S.-L.}\ \bibnamefont {Yin}}, \ and\ \bibinfo
  {author} {\bibfnamefont {D.-M.}\ \bibnamefont {Chen}},\ }\bibfield  {title}
  {\enquote {\bibinfo {title} {Origin of the metallic to insulating transition
  of an epitaxial {Bi}\hkl(1 1 1) film grown on {Si}\hkl(1 1 1)},}\ }\href@noop
  {} {\bibfield  {journal} {\bibinfo  {journal} {Chin. Phys. B}\ }\textbf
  {\bibinfo {volume} {19}},\ \bibinfo {pages} {087201} (\bibinfo {year}
  {2010})}\BibitemShut {NoStop}%
\bibitem [{\citenamefont {Xiao}\ \emph {et~al.}(2012)\citenamefont {Xiao},
  \citenamefont {Wei},\ and\ \citenamefont {Jin}}]{xiao2012}%
  \BibitemOpen
  \bibfield  {author} {\bibinfo {author} {\bibfnamefont {S.}~\bibnamefont
  {Xiao}}, \bibinfo {author} {\bibfnamefont {D.}~\bibnamefont {Wei}}, \ and\
  \bibinfo {author} {\bibfnamefont {X.}~\bibnamefont {Jin}},\ }\bibfield
  {title} {\enquote {\bibinfo {title} {{Bi}\hkl(111) thin film with insulating
  interior but metallic surfaces},}\ }\href@noop {} {\bibfield  {journal}
  {\bibinfo  {journal} {Phys. Rev. Lett.}\ }\textbf {\bibinfo {volume} {109}},\
  \bibinfo {pages} {166805} (\bibinfo {year} {2012})}\BibitemShut {NoStop}%
\bibitem [{\citenamefont {Zhu}\ \emph {et~al.}(2016)\citenamefont {Zhu},
  \citenamefont {Wu}, \citenamefont {Gong}, \citenamefont {Xiao},\ and\
  \citenamefont {Jin}}]{zhu2016}%
  \BibitemOpen
  \bibfield  {author} {\bibinfo {author} {\bibfnamefont {K.}~\bibnamefont
  {Zhu}}, \bibinfo {author} {\bibfnamefont {L.}~\bibnamefont {Wu}}, \bibinfo
  {author} {\bibfnamefont {X.}~\bibnamefont {Gong}}, \bibinfo {author}
  {\bibfnamefont {S.}~\bibnamefont {Xiao}}, \ and\ \bibinfo {author}
  {\bibfnamefont {X.}~\bibnamefont {Jin}},\ }\bibfield  {title} {\enquote
  {\bibinfo {title} {Quantum transport in the surface states of epitaxial
  {Bi}\hkl(1 1 1) thin films},}\ }\href@noop {} {\bibfield  {journal} {\bibinfo
   {journal} {Phys. Rev. B}\ }\textbf {\bibinfo {volume} {94}},\ \bibinfo
  {pages} {121401(R)} (\bibinfo {year} {2016})}\BibitemShut {NoStop}%
\bibitem [{\citenamefont {Hirahara}\ and\ \citenamefont
  {Hasegawa}(2018)}]{hirahara2018}%
  \BibitemOpen
  \bibfield  {author} {\bibinfo {author} {\bibfnamefont {T.}~\bibnamefont
  {Hirahara}}\ and\ \bibinfo {author} {\bibfnamefont {S.}~\bibnamefont
  {Hasegawa}},\ }\bibfield  {title} {\enquote {\bibinfo {title} {Comment on
  "{Quantum} transport in the surface states of epitaxial {Bi}\hkl(1 1 1) thin
  films"},}\ }\href@noop {} {\bibfield  {journal} {\bibinfo  {journal} {Phys.
  Rev. B}\ }\textbf {\bibinfo {volume} {97}},\ \bibinfo {pages} {207401}
  (\bibinfo {year} {2018})}\BibitemShut {NoStop}%
\bibitem [{\citenamefont {Zhu}\ \emph {et~al.}(2018)\citenamefont {Zhu},
  \citenamefont {Wu}, \citenamefont {Gong}, \citenamefont {Xiao},\ and\
  \citenamefont {Jin}}]{zhu2018}%
  \BibitemOpen
  \bibfield  {author} {\bibinfo {author} {\bibfnamefont {K.}~\bibnamefont
  {Zhu}}, \bibinfo {author} {\bibfnamefont {L.}~\bibnamefont {Wu}}, \bibinfo
  {author} {\bibfnamefont {X.}~\bibnamefont {Gong}}, \bibinfo {author}
  {\bibfnamefont {S.}~\bibnamefont {Xiao}}, \ and\ \bibinfo {author}
  {\bibfnamefont {X.}~\bibnamefont {Jin}},\ }\bibfield  {title} {\enquote
  {\bibinfo {title} {Reply to "{Comment} on '{Quantum} transport in the surface
  states of epitaxial {Bi}\hkl(1 1 1) thin films'"},}\ }\href@noop {}
  {\bibfield  {journal} {\bibinfo  {journal} {Phys. Rev. B}\ }\textbf {\bibinfo
  {volume} {97}},\ \bibinfo {pages} {207402} (\bibinfo {year}
  {2018})}\BibitemShut {NoStop}%
\bibitem [{\citenamefont {Kröger}\ \emph {et~al.}(2018)\citenamefont
  {Kröger}, \citenamefont {Abdelbarey}, \citenamefont {Siemens}, \citenamefont
  {Lükermann}, \citenamefont {Sologub}, \citenamefont {Pfnür},\ and\
  \citenamefont {Tegenkamp}}]{kroeger2018}%
  \BibitemOpen
  \bibfield  {author} {\bibinfo {author} {\bibfnamefont {P.}~\bibnamefont
  {Kröger}}, \bibinfo {author} {\bibfnamefont {D.}~\bibnamefont {Abdelbarey}},
  \bibinfo {author} {\bibfnamefont {M.}~\bibnamefont {Siemens}}, \bibinfo
  {author} {\bibfnamefont {D.}~\bibnamefont {Lükermann}}, \bibinfo {author}
  {\bibfnamefont {S.}~\bibnamefont {Sologub}}, \bibinfo {author} {\bibfnamefont
  {H.}~\bibnamefont {Pfnür}}, \ and\ \bibinfo {author} {\bibfnamefont
  {C.}~\bibnamefont {Tegenkamp}},\ }\bibfield  {title} {\enquote {\bibinfo
  {title} {Controlling conductivity by quantum well states in ultrathin
  {Bi}\hkl(111) films},}\ }\href@noop {} {\bibfield  {journal} {\bibinfo
  {journal} {Phys. Rev. B}\ }\textbf {\bibinfo {volume} {97}},\ \bibinfo
  {pages} {045403} (\bibinfo {year} {2018})}\BibitemShut {NoStop}%
\bibitem [{\citenamefont {Lükermann}\ \emph {et~al.}(2013)\citenamefont
  {Lükermann}, \citenamefont {Sologub}, \citenamefont {Pfnür}, \citenamefont
  {Klein}, \citenamefont {{Horn-von-Hoegen}},\ and\ \citenamefont
  {Tegenkamp}}]{luekermann2013}%
  \BibitemOpen
  \bibfield  {author} {\bibinfo {author} {\bibfnamefont {D.}~\bibnamefont
  {Lükermann}}, \bibinfo {author} {\bibfnamefont {S.}~\bibnamefont {Sologub}},
  \bibinfo {author} {\bibfnamefont {H.}~\bibnamefont {Pfnür}}, \bibinfo
  {author} {\bibfnamefont {C.}~\bibnamefont {Klein}}, \bibinfo {author}
  {\bibfnamefont {M.}~\bibnamefont {{Horn-von-Hoegen}}}, \ and\ \bibinfo
  {author} {\bibfnamefont {C.}~\bibnamefont {Tegenkamp}},\ }\bibfield  {title}
  {\enquote {\bibinfo {title} {Effect of adsorbed magnetic and non-magnetic
  atoms on electronic transport through surfaces with strong spin-orbit
  coupling},}\ }\href@noop {} {\bibfield  {journal} {\bibinfo  {journal}
  {Mat.-wiss. u. Werkstofftech.}\ }\textbf {\bibinfo {volume} {44}},\ \bibinfo
  {pages} {210} (\bibinfo {year} {2013})}\BibitemShut {NoStop}%
\bibitem [{\citenamefont {Aitani}\ \emph {et~al.}(2014)\citenamefont {Aitani},
  \citenamefont {Hirahara}, \citenamefont {Ichinokura}, \citenamefont
  {Hanaduka}, \citenamefont {Shin},\ and\ \citenamefont
  {Hasegawa}}]{aitani2014}%
  \BibitemOpen
  \bibfield  {author} {\bibinfo {author} {\bibfnamefont {M.}~\bibnamefont
  {Aitani}}, \bibinfo {author} {\bibfnamefont {T.}~\bibnamefont {Hirahara}},
  \bibinfo {author} {\bibfnamefont {S.}~\bibnamefont {Ichinokura}}, \bibinfo
  {author} {\bibfnamefont {M.}~\bibnamefont {Hanaduka}}, \bibinfo {author}
  {\bibfnamefont {D.}~\bibnamefont {Shin}}, \ and\ \bibinfo {author}
  {\bibfnamefont {S.}~\bibnamefont {Hasegawa}},\ }\bibfield  {title} {\enquote
  {\bibinfo {title} {In situ magnetotransport measurements in ultrathin {Bi}
  films: {E}vidence for surface-bulk coherent transport},}\ }\href@noop {}
  {\bibfield  {journal} {\bibinfo  {journal} {Phys. Rev. Lett.}\ }\textbf
  {\bibinfo {volume} {113}},\ \bibinfo {pages} {206802} (\bibinfo {year}
  {2014})}\BibitemShut {NoStop}%
\bibitem [{\citenamefont {Abdelbarey}\ \emph {et~al.}(2020)\citenamefont
  {Abdelbarey}, \citenamefont {Koch}, \citenamefont {Mamiyev}, \citenamefont
  {Tegenkamp},\ and\ \citenamefont {Pfnür}}]{abdelbarey2020}%
  \BibitemOpen
  \bibfield  {author} {\bibinfo {author} {\bibfnamefont {D.}~\bibnamefont
  {Abdelbarey}}, \bibinfo {author} {\bibfnamefont {J.}~\bibnamefont {Koch}},
  \bibinfo {author} {\bibfnamefont {Z.}~\bibnamefont {Mamiyev}}, \bibinfo
  {author} {\bibfnamefont {C.}~\bibnamefont {Tegenkamp}}, \ and\ \bibinfo
  {author} {\bibfnamefont {H.}~\bibnamefont {Pfnür}},\ }\bibfield  {title}
  {\enquote {\bibinfo {title} {Thickness-dependent electronic transport through
  epitaxial nontrivial {Bi} quantum films},}\ }\href@noop {} {\bibfield
  {journal} {\bibinfo  {journal} {Phys. Rev. B}\ }\textbf {\bibinfo {volume}
  {102}},\ \bibinfo {pages} {115409} (\bibinfo {year} {2020})}\BibitemShut
  {NoStop}%
\bibitem [{\citenamefont {Chang}\ \emph {et~al.}(2019)\citenamefont {Chang},
  \citenamefont {Lu}, \citenamefont {Wang}, \citenamefont {Lin}, \citenamefont
  {Miller}, \citenamefont {Chiang},\ and\ \citenamefont {Bian}}]{chang2019}%
  \BibitemOpen
  \bibfield  {author} {\bibinfo {author} {\bibfnamefont {T.-R.}\ \bibnamefont
  {Chang}}, \bibinfo {author} {\bibfnamefont {Q.}~\bibnamefont {Lu}}, \bibinfo
  {author} {\bibfnamefont {X.}~\bibnamefont {Wang}}, \bibinfo {author}
  {\bibfnamefont {H.}~\bibnamefont {Lin}}, \bibinfo {author} {\bibfnamefont
  {T.}~\bibnamefont {Miller}}, \bibinfo {author} {\bibfnamefont {T.-C.}\
  \bibnamefont {Chiang}}, \ and\ \bibinfo {author} {\bibfnamefont
  {G.}~\bibnamefont {Bian}},\ }\bibfield  {title} {\enquote {\bibinfo {title}
  {Band topology of bismuth quantum films},}\ }\href@noop {} {\bibfield
  {journal} {\bibinfo  {journal} {Crystals}\ }\textbf {\bibinfo {volume} {9}},\
  \bibinfo {pages} {510} (\bibinfo {year} {2019})}\BibitemShut {NoStop}%
\bibitem [{\citenamefont {Hasan}\ and\ \citenamefont {Kane}(2010)}]{hasan2010}%
  \BibitemOpen
  \bibfield  {author} {\bibinfo {author} {\bibfnamefont {M.~Z.}\ \bibnamefont
  {Hasan}}\ and\ \bibinfo {author} {\bibfnamefont {C.~L.}\ \bibnamefont
  {Kane}},\ }\bibfield  {title} {\enquote {\bibinfo {title} {Colloquium:
  {Topological} insulators},}\ }\href@noop {} {\bibfield  {journal} {\bibinfo
  {journal} {Rev. Mod. Phys.}\ }\textbf {\bibinfo {volume} {82}},\ \bibinfo
  {pages} {3045} (\bibinfo {year} {2010})}\BibitemShut {NoStop}%
\bibitem [{\citenamefont {Liu}\ and\ \citenamefont {Allen}(1995)}]{liu1995}%
  \BibitemOpen
  \bibfield  {author} {\bibinfo {author} {\bibfnamefont {Y.}~\bibnamefont
  {Liu}}\ and\ \bibinfo {author} {\bibfnamefont {R.~E.}\ \bibnamefont
  {Allen}},\ }\bibfield  {title} {\enquote {\bibinfo {title} {Electronic
  structure of the semimetals {Bi} and {Sb}},}\ }\href@noop {} {\bibfield
  {journal} {\bibinfo  {journal} {Phys. Rev. B}\ }\textbf {\bibinfo {volume}
  {52}},\ \bibinfo {pages} {1566} (\bibinfo {year} {1995})}\BibitemShut
  {NoStop}%
\bibitem [{\citenamefont {Fukui}\ and\ \citenamefont
  {Hatsugai}(2007)}]{fukui2007}%
  \BibitemOpen
  \bibfield  {author} {\bibinfo {author} {\bibfnamefont {T.}~\bibnamefont
  {Fukui}}\ and\ \bibinfo {author} {\bibfnamefont {Y.}~\bibnamefont
  {Hatsugai}},\ }\bibfield  {title} {\enquote {\bibinfo {title} {Quantum {S}pin
  {H}all effect in three dimensional materials: Lattice computation of {Z}$_2$
  topological invariants and its application to {Bi} and {Sb}},}\ }\href@noop
  {} {\bibfield  {journal} {\bibinfo  {journal} {J. Phys. Soc. Jpn.}\ }\textbf
  {\bibinfo {volume} {76}},\ \bibinfo {pages} {053702} (\bibinfo {year}
  {2007})}\BibitemShut {NoStop}%
\bibitem [{\citenamefont {Ohtsubo}\ \emph {et~al.}(2013)\citenamefont
  {Ohtsubo}, \citenamefont {Perfetti}, \citenamefont {Goerbig}, \citenamefont
  {Le~F{\`{e}}vre}, \citenamefont {Bertran},\ and\ \citenamefont
  {Taleb-Ibrahimi}}]{ohtsubo2013}%
  \BibitemOpen
  \bibfield  {author} {\bibinfo {author} {\bibfnamefont {Y.}~\bibnamefont
  {Ohtsubo}}, \bibinfo {author} {\bibfnamefont {L.}~\bibnamefont {Perfetti}},
  \bibinfo {author} {\bibfnamefont {M.~O.}\ \bibnamefont {Goerbig}}, \bibinfo
  {author} {\bibfnamefont {P.}~\bibnamefont {Le~F{\`{e}}vre}}, \bibinfo
  {author} {\bibfnamefont {F.}~\bibnamefont {Bertran}}, \ and\ \bibinfo
  {author} {\bibfnamefont {A.}~\bibnamefont {Taleb-Ibrahimi}},\ }\bibfield
  {title} {\enquote {\bibinfo {title} {Non-trivial surface-band dispersion on
  {Bi}\hkl(111)},}\ }\href@noop {} {\bibfield  {journal} {\bibinfo  {journal}
  {New J. Phys.}\ }\textbf {\bibinfo {volume} {15}},\ \bibinfo {pages} {033041}
  (\bibinfo {year} {2013})}\BibitemShut {NoStop}%
\bibitem [{\citenamefont {Ohtsubo}\ and\ \citenamefont
  {Kimura}(2016)}]{ohtsubo2016}%
  \BibitemOpen
  \bibfield  {author} {\bibinfo {author} {\bibfnamefont {Y.}~\bibnamefont
  {Ohtsubo}}\ and\ \bibinfo {author} {\bibfnamefont {S.}~\bibnamefont
  {Kimura}},\ }\bibfield  {title} {\enquote {\bibinfo {title} {Topological
  phase transition of single-crystal {Bi} based on empirical tight-binding
  calculations},}\ }\href@noop {} {\bibfield  {journal} {\bibinfo  {journal}
  {New J. Phys.}\ }\textbf {\bibinfo {volume} {18}},\ \bibinfo {pages} {123015}
  (\bibinfo {year} {2016})}\BibitemShut {NoStop}%
\bibitem [{\citenamefont {Saito}\ \emph {et~al.}(2016)\citenamefont {Saito},
  \citenamefont {Sawahata}, \citenamefont {Komine},\ and\ \citenamefont
  {Aono}}]{saito2016}%
  \BibitemOpen
  \bibfield  {author} {\bibinfo {author} {\bibfnamefont {K.}~\bibnamefont
  {Saito}}, \bibinfo {author} {\bibfnamefont {H.}~\bibnamefont {Sawahata}},
  \bibinfo {author} {\bibfnamefont {T.}~\bibnamefont {Komine}}, \ and\ \bibinfo
  {author} {\bibfnamefont {T.}~\bibnamefont {Aono}},\ }\bibfield  {title}
  {\enquote {\bibinfo {title} {Tight-binding theory of surface spin states on
  bismuth thin films},}\ }\href@noop {} {\bibfield  {journal} {\bibinfo
  {journal} {Phys. Rev. B}\ }\textbf {\bibinfo {volume} {93}},\ \bibinfo
  {pages} {041301(R)} (\bibinfo {year} {2016})}\BibitemShut {NoStop}%
\bibitem [{\citenamefont {Aryasetiawan}\ and\ \citenamefont
  {Gunnarsson}(1998)}]{aryasetiawan1998}%
  \BibitemOpen
  \bibfield  {author} {\bibinfo {author} {\bibfnamefont {F.}~\bibnamefont
  {Aryasetiawan}}\ and\ \bibinfo {author} {\bibfnamefont {O.}~\bibnamefont
  {Gunnarsson}},\ }\bibfield  {title} {\enquote {\bibinfo {title} {The {GW}
  method},}\ }\href@noop {} {\bibfield  {journal} {\bibinfo  {journal} {Rep.
  Prog. Phys.}\ }\textbf {\bibinfo {volume} {61}},\ \bibinfo {pages} {237}
  (\bibinfo {year} {1998})}\BibitemShut {NoStop}%
\bibitem [{\citenamefont {Almbladh}\ and\ \citenamefont {von
  Barth}(1985)}]{almbladh1985}%
  \BibitemOpen
  \bibfield  {author} {\bibinfo {author} {\bibfnamefont {C.-O.}\ \bibnamefont
  {Almbladh}}\ and\ \bibinfo {author} {\bibfnamefont {U.}~\bibnamefont {von
  Barth}},\ }\bibfield  {title} {\enquote {\bibinfo {title} {Exact results for
  the charge and spin densities, exchange-correlation potentials, and
  density-functional eigenvalues},}\ }\href@noop {} {\bibfield  {journal}
  {\bibinfo  {journal} {Phys. Rev. B}\ }\textbf {\bibinfo {volume} {31}},\
  \bibinfo {pages} {3231} (\bibinfo {year} {1985})}\BibitemShut {NoStop}%
\bibitem [{\citenamefont {Giannozzi}\ \emph {et~al.}(2009)\citenamefont
  {Giannozzi}, \citenamefont {Baroni}, \citenamefont {Bonini}, \citenamefont
  {Calandra}, \citenamefont {Car}, \citenamefont {Cavazzoni}, \citenamefont
  {Ceresoli}, \citenamefont {Chiarotti}, \citenamefont {Cococcioni},
  \citenamefont {Dabo}, \citenamefont {Dal~Corso}, \citenamefont
  {de~Gironcoli}, \citenamefont {Fabris}, \citenamefont {Fratesi},
  \citenamefont {Gebauer}, \citenamefont {Gerstmann}, \citenamefont
  {Gougoussis}, \citenamefont {Kokalj}, \citenamefont {Lazzeri}, \citenamefont
  {Martin-Samos}, \citenamefont {Marzari}, \citenamefont {Mauri}, \citenamefont
  {Mazzarello}, \citenamefont {Paolini}, \citenamefont {Pasquarello},
  \citenamefont {Paulatto}, \citenamefont {Sbraccia}, \citenamefont {Scandolo},
  \citenamefont {Sclauzero}, \citenamefont {Seitsonen}, \citenamefont
  {Smogunov}, \citenamefont {Umari},\ and\ \citenamefont
  {Wentzcovitch}}]{giannozzi2009}%
  \BibitemOpen
  \bibfield  {author} {\bibinfo {author} {\bibfnamefont {P.}~\bibnamefont
  {Giannozzi}}, \bibinfo {author} {\bibfnamefont {S.}~\bibnamefont {Baroni}},
  \bibinfo {author} {\bibfnamefont {N.}~\bibnamefont {Bonini}}, \bibinfo
  {author} {\bibfnamefont {M.}~\bibnamefont {Calandra}}, \bibinfo {author}
  {\bibfnamefont {R.}~\bibnamefont {Car}}, \bibinfo {author} {\bibfnamefont
  {C.}~\bibnamefont {Cavazzoni}}, \bibinfo {author} {\bibfnamefont
  {D.}~\bibnamefont {Ceresoli}}, \bibinfo {author} {\bibfnamefont {G.~L.}\
  \bibnamefont {Chiarotti}}, \bibinfo {author} {\bibfnamefont {M.}~\bibnamefont
  {Cococcioni}}, \bibinfo {author} {\bibfnamefont {I.}~\bibnamefont {Dabo}},
  \bibinfo {author} {\bibfnamefont {A.}~\bibnamefont {Dal~Corso}}, \bibinfo
  {author} {\bibfnamefont {S.}~\bibnamefont {de~Gironcoli}}, \bibinfo {author}
  {\bibfnamefont {S.}~\bibnamefont {Fabris}}, \bibinfo {author} {\bibfnamefont
  {G.}~\bibnamefont {Fratesi}}, \bibinfo {author} {\bibfnamefont
  {R.}~\bibnamefont {Gebauer}}, \bibinfo {author} {\bibfnamefont
  {U.}~\bibnamefont {Gerstmann}}, \bibinfo {author} {\bibfnamefont
  {C.}~\bibnamefont {Gougoussis}}, \bibinfo {author} {\bibfnamefont
  {A.}~\bibnamefont {Kokalj}}, \bibinfo {author} {\bibfnamefont
  {M.}~\bibnamefont {Lazzeri}}, \bibinfo {author} {\bibfnamefont
  {L.}~\bibnamefont {Martin-Samos}}, \bibinfo {author} {\bibfnamefont
  {N.}~\bibnamefont {Marzari}}, \bibinfo {author} {\bibfnamefont
  {F.}~\bibnamefont {Mauri}}, \bibinfo {author} {\bibfnamefont
  {R.}~\bibnamefont {Mazzarello}}, \bibinfo {author} {\bibfnamefont
  {S.}~\bibnamefont {Paolini}}, \bibinfo {author} {\bibfnamefont
  {A.}~\bibnamefont {Pasquarello}}, \bibinfo {author} {\bibfnamefont
  {L.}~\bibnamefont {Paulatto}}, \bibinfo {author} {\bibfnamefont
  {C.}~\bibnamefont {Sbraccia}}, \bibinfo {author} {\bibfnamefont
  {S.}~\bibnamefont {Scandolo}}, \bibinfo {author} {\bibfnamefont
  {G.}~\bibnamefont {Sclauzero}}, \bibinfo {author} {\bibfnamefont {A.~P.}\
  \bibnamefont {Seitsonen}}, \bibinfo {author} {\bibfnamefont {A.}~\bibnamefont
  {Smogunov}}, \bibinfo {author} {\bibfnamefont {P.}~\bibnamefont {Umari}}, \
  and\ \bibinfo {author} {\bibfnamefont {R.~M.}\ \bibnamefont {Wentzcovitch}},\
  }\bibfield  {title} {\enquote {\bibinfo {title} {\textsc{Quantum Espresso}: A
  modular and open-source software project for quantum simulations of
  materials},}\ }\href@noop {} {\bibfield  {journal} {\bibinfo  {journal} {J.
  Phys.: Condens. Matter}\ }\textbf {\bibinfo {volume} {21}},\ \bibinfo {pages}
  {395502} (\bibinfo {year} {2009})}\BibitemShut {NoStop}%
\bibitem [{\citenamefont {Giannozzi}\ \emph {et~al.}(2017)\citenamefont
  {Giannozzi}, \citenamefont {Andreussi}, \citenamefont {Brumme}, \citenamefont
  {Bunau}, \citenamefont {Buongiorno~Nardelli}, \citenamefont {Calandra},
  \citenamefont {Car}, \citenamefont {Cavazzoni}, \citenamefont {Ceresoli},
  \citenamefont {Cococcioni}, \citenamefont {Colonna}, \citenamefont
  {Carnimeo}, \citenamefont {Dal~Corso}, \citenamefont {de~Gironcoli},
  \citenamefont {Delugas}, \citenamefont {DiStasio~Jr.}, \citenamefont
  {Ferretti}, \citenamefont {Floris}, \citenamefont {Fratesi}, \citenamefont
  {Fugallo}, \citenamefont {Gebauer}, \citenamefont {Gerstmann}, \citenamefont
  {Giustino}, \citenamefont {Gorni}, \citenamefont {Jia}, \citenamefont
  {Kawamura}, \citenamefont {Ko}, \citenamefont {Kokalj}, \citenamefont
  {Kü{\c{c}}ükbenli}, \citenamefont {Lazzeri}, \citenamefont {Marsili},
  \citenamefont {Marzari}, \citenamefont {Mauri}, \citenamefont {Nguyen},
  \citenamefont {Nguyen}, \citenamefont {{Otero-de-la-Roza}}, \citenamefont
  {Paulatto}, \citenamefont {Ponc{\'{e}}}, \citenamefont {Rocca}, \citenamefont
  {Sabatini}, \citenamefont {Santra}, \citenamefont {Schlipf}, \citenamefont
  {Seitsonen}, \citenamefont {Smogunov}, \citenamefont {Timrov}, \citenamefont
  {Thonhauser}, \citenamefont {Umari}, \citenamefont {Vast}, \citenamefont
  {Wu},\ and\ \citenamefont {Baroni}}]{giannozzi2017}%
  \BibitemOpen
  \bibfield  {author} {\bibinfo {author} {\bibfnamefont {P.}~\bibnamefont
  {Giannozzi}}, \bibinfo {author} {\bibfnamefont {O.}~\bibnamefont
  {Andreussi}}, \bibinfo {author} {\bibfnamefont {T.}~\bibnamefont {Brumme}},
  \bibinfo {author} {\bibfnamefont {O.}~\bibnamefont {Bunau}}, \bibinfo
  {author} {\bibfnamefont {M.}~\bibnamefont {Buongiorno~Nardelli}}, \bibinfo
  {author} {\bibfnamefont {M.}~\bibnamefont {Calandra}}, \bibinfo {author}
  {\bibfnamefont {R.}~\bibnamefont {Car}}, \bibinfo {author} {\bibfnamefont
  {C.}~\bibnamefont {Cavazzoni}}, \bibinfo {author} {\bibfnamefont
  {D.}~\bibnamefont {Ceresoli}}, \bibinfo {author} {\bibfnamefont
  {M.}~\bibnamefont {Cococcioni}}, \bibinfo {author} {\bibfnamefont
  {N.}~\bibnamefont {Colonna}}, \bibinfo {author} {\bibfnamefont
  {I.}~\bibnamefont {Carnimeo}}, \bibinfo {author} {\bibfnamefont
  {A.}~\bibnamefont {Dal~Corso}}, \bibinfo {author} {\bibfnamefont
  {S.}~\bibnamefont {de~Gironcoli}}, \bibinfo {author} {\bibfnamefont
  {P.}~\bibnamefont {Delugas}}, \bibinfo {author} {\bibfnamefont {R.~A.}\
  \bibnamefont {DiStasio~Jr.}}, \bibinfo {author} {\bibfnamefont
  {A.}~\bibnamefont {Ferretti}}, \bibinfo {author} {\bibfnamefont
  {A.}~\bibnamefont {Floris}}, \bibinfo {author} {\bibfnamefont
  {G.}~\bibnamefont {Fratesi}}, \bibinfo {author} {\bibfnamefont
  {G.}~\bibnamefont {Fugallo}}, \bibinfo {author} {\bibfnamefont
  {R.}~\bibnamefont {Gebauer}}, \bibinfo {author} {\bibfnamefont
  {U.}~\bibnamefont {Gerstmann}}, \bibinfo {author} {\bibfnamefont
  {F.}~\bibnamefont {Giustino}}, \bibinfo {author} {\bibfnamefont
  {T.}~\bibnamefont {Gorni}}, \bibinfo {author} {\bibfnamefont
  {J.}~\bibnamefont {Jia}}, \bibinfo {author} {\bibfnamefont {M.}~\bibnamefont
  {Kawamura}}, \bibinfo {author} {\bibfnamefont {H.-Y.}\ \bibnamefont {Ko}},
  \bibinfo {author} {\bibfnamefont {A.}~\bibnamefont {Kokalj}}, \bibinfo
  {author} {\bibfnamefont {E.}~\bibnamefont {Kü{\c{c}}ükbenli}}, \bibinfo
  {author} {\bibfnamefont {M.}~\bibnamefont {Lazzeri}}, \bibinfo {author}
  {\bibfnamefont {M.}~\bibnamefont {Marsili}}, \bibinfo {author} {\bibfnamefont
  {N.}~\bibnamefont {Marzari}}, \bibinfo {author} {\bibfnamefont
  {F.}~\bibnamefont {Mauri}}, \bibinfo {author} {\bibfnamefont {N.~L.}\
  \bibnamefont {Nguyen}}, \bibinfo {author} {\bibfnamefont {H.-V.}\
  \bibnamefont {Nguyen}}, \bibinfo {author} {\bibfnamefont {A.}~\bibnamefont
  {{Otero-de-la-Roza}}}, \bibinfo {author} {\bibfnamefont {L.}~\bibnamefont
  {Paulatto}}, \bibinfo {author} {\bibfnamefont {S.}~\bibnamefont
  {Ponc{\'{e}}}}, \bibinfo {author} {\bibfnamefont {D.}~\bibnamefont {Rocca}},
  \bibinfo {author} {\bibfnamefont {R.}~\bibnamefont {Sabatini}}, \bibinfo
  {author} {\bibfnamefont {B.}~\bibnamefont {Santra}}, \bibinfo {author}
  {\bibfnamefont {M.}~\bibnamefont {Schlipf}}, \bibinfo {author} {\bibfnamefont
  {A.~P.}\ \bibnamefont {Seitsonen}}, \bibinfo {author} {\bibfnamefont
  {A.}~\bibnamefont {Smogunov}}, \bibinfo {author} {\bibfnamefont
  {I.}~\bibnamefont {Timrov}}, \bibinfo {author} {\bibfnamefont
  {T.}~\bibnamefont {Thonhauser}}, \bibinfo {author} {\bibfnamefont
  {P.}~\bibnamefont {Umari}}, \bibinfo {author} {\bibfnamefont
  {N.}~\bibnamefont {Vast}}, \bibinfo {author} {\bibfnamefont {X.}~\bibnamefont
  {Wu}}, \ and\ \bibinfo {author} {\bibfnamefont {S.}~\bibnamefont {Baroni}},\
  }\bibfield  {title} {\enquote {\bibinfo {title} {Advanced capabilities for
  materials modelling with \textsc{Quantum Espresso}},}\ }\href@noop {}
  {\bibfield  {journal} {\bibinfo  {journal} {J. Phys.: Condens. Matter}\
  }\textbf {\bibinfo {volume} {29}},\ \bibinfo {pages} {465901} (\bibinfo
  {year} {2017})}\BibitemShut {NoStop}%
\bibitem [{\citenamefont {Perdew}\ \emph {et~al.}(1996)\citenamefont {Perdew},
  \citenamefont {Burke},\ and\ \citenamefont {Ernzerhof}}]{perdew1996}%
  \BibitemOpen
  \bibfield  {author} {\bibinfo {author} {\bibfnamefont {J.~P.}\ \bibnamefont
  {Perdew}}, \bibinfo {author} {\bibfnamefont {K.}~\bibnamefont {Burke}}, \
  and\ \bibinfo {author} {\bibfnamefont {M.}~\bibnamefont {Ernzerhof}},\
  }\bibfield  {title} {\enquote {\bibinfo {title} {Generalized gradient
  approximation made simple},}\ }\href@noop {} {\bibfield  {journal} {\bibinfo
  {journal} {Phys. Rev. Lett.}\ }\textbf {\bibinfo {volume} {77}},\ \bibinfo
  {pages} {3865} (\bibinfo {year} {1996})}\BibitemShut {NoStop}%
\bibitem [{\citenamefont {Hamann}(2013)}]{hamann2013}%
  \BibitemOpen
  \bibfield  {author} {\bibinfo {author} {\bibfnamefont {D.~R.}\ \bibnamefont
  {Hamann}},\ }\bibfield  {title} {\enquote {\bibinfo {title} {Optimized
  norm-conserving {Vanderbilt} pseudopotentials},}\ }\href@noop {} {\bibfield
  {journal} {\bibinfo  {journal} {Phys. Rev. B}\ }\textbf {\bibinfo {volume}
  {88}},\ \bibinfo {pages} {085117} (\bibinfo {year} {2013})}\BibitemShut
  {NoStop}%
\bibitem [{\citenamefont {Schlipf}\ and\ \citenamefont
  {Gygi}(2015)}]{schlipf2015}%
  \BibitemOpen
  \bibfield  {author} {\bibinfo {author} {\bibfnamefont {M.}~\bibnamefont
  {Schlipf}}\ and\ \bibinfo {author} {\bibfnamefont {F.}~\bibnamefont {Gygi}},\
  }\bibfield  {title} {\enquote {\bibinfo {title} {Optimization algorithm for
  the generation of {ONCV} pseudopotentials},}\ }\href@noop {} {\bibfield
  {journal} {\bibinfo  {journal} {Comput. Phys. Commun.}\ }\textbf {\bibinfo
  {volume} {196}},\ \bibinfo {pages} {36} (\bibinfo {year} {2015})}\BibitemShut
  {NoStop}%
\bibitem [{\citenamefont {Scherpelz}\ \emph {et~al.}(2016)\citenamefont
  {Scherpelz}, \citenamefont {Govoni}, \citenamefont {Hamada},\ and\
  \citenamefont {Galli}}]{scherpelz2016}%
  \BibitemOpen
  \bibfield  {author} {\bibinfo {author} {\bibfnamefont {P.}~\bibnamefont
  {Scherpelz}}, \bibinfo {author} {\bibfnamefont {M.}~\bibnamefont {Govoni}},
  \bibinfo {author} {\bibfnamefont {I.}~\bibnamefont {Hamada}}, \ and\ \bibinfo
  {author} {\bibfnamefont {G.}~\bibnamefont {Galli}},\ }\bibfield  {title}
  {\enquote {\bibinfo {title} {Implementation and validation of fully
  relativistic {GW} calculations: {Spin}-orbit coupling in molecules,
  nanocrystals, and solids},}\ }\href@noop {} {\bibfield  {journal} {\bibinfo
  {journal} {J. Chem. Theory Comput.}\ }\textbf {\bibinfo {volume} {12}},\
  \bibinfo {pages} {3523} (\bibinfo {year} {2016})}\BibitemShut {NoStop}%
\bibitem [{\citenamefont {Marini}\ \emph {et~al.}(2009)\citenamefont {Marini},
  \citenamefont {Hogan}, \citenamefont {Grüning},\ and\ \citenamefont
  {Varsano}}]{marini2009}%
  \BibitemOpen
  \bibfield  {author} {\bibinfo {author} {\bibfnamefont {A.}~\bibnamefont
  {Marini}}, \bibinfo {author} {\bibfnamefont {C.}~\bibnamefont {Hogan}},
  \bibinfo {author} {\bibfnamefont {M.}~\bibnamefont {Grüning}}, \ and\
  \bibinfo {author} {\bibfnamefont {D.}~\bibnamefont {Varsano}},\ }\bibfield
  {title} {\enquote {\bibinfo {title} {\textsc{Yambo}: An ab initio tool for
  excited state calculations},}\ }\href@noop {} {\bibfield  {journal} {\bibinfo
   {journal} {Comput. Phys. Commun.}\ }\textbf {\bibinfo {volume} {180}},\
  \bibinfo {pages} {1392} (\bibinfo {year} {2009})}\BibitemShut {NoStop}%
\bibitem [{\citenamefont {Sangalli}\ \emph {et~al.}(2019)\citenamefont
  {Sangalli}, \citenamefont {Ferretti}, \citenamefont {Miranda}, \citenamefont
  {Attaccalite}, \citenamefont {Marri}, \citenamefont {Cannuccia},
  \citenamefont {Melo}, \citenamefont {Marsili}, \citenamefont {Paleari},
  \citenamefont {Marrazzo}, \citenamefont {Prandini}, \citenamefont
  {Bonf{\`{a}}}, \citenamefont {Atambo}, \citenamefont {Affinito},
  \citenamefont {Palummo}, \citenamefont {Molina-S{\'{a}}nchez}, \citenamefont
  {Hogan}, \citenamefont {Grüning}, \citenamefont {Varsano},\ and\
  \citenamefont {Marini}}]{sangalli2019}%
  \BibitemOpen
  \bibfield  {author} {\bibinfo {author} {\bibfnamefont {D.}~\bibnamefont
  {Sangalli}}, \bibinfo {author} {\bibfnamefont {A.}~\bibnamefont {Ferretti}},
  \bibinfo {author} {\bibfnamefont {H.}~\bibnamefont {Miranda}}, \bibinfo
  {author} {\bibfnamefont {C.}~\bibnamefont {Attaccalite}}, \bibinfo {author}
  {\bibfnamefont {I.}~\bibnamefont {Marri}}, \bibinfo {author} {\bibfnamefont
  {E.}~\bibnamefont {Cannuccia}}, \bibinfo {author} {\bibfnamefont
  {P.}~\bibnamefont {Melo}}, \bibinfo {author} {\bibfnamefont {M.}~\bibnamefont
  {Marsili}}, \bibinfo {author} {\bibfnamefont {F.}~\bibnamefont {Paleari}},
  \bibinfo {author} {\bibfnamefont {A.}~\bibnamefont {Marrazzo}}, \bibinfo
  {author} {\bibfnamefont {G.}~\bibnamefont {Prandini}}, \bibinfo {author}
  {\bibfnamefont {P.}~\bibnamefont {Bonf{\`{a}}}}, \bibinfo {author}
  {\bibfnamefont {M.~O.}\ \bibnamefont {Atambo}}, \bibinfo {author}
  {\bibfnamefont {F.}~\bibnamefont {Affinito}}, \bibinfo {author}
  {\bibfnamefont {M.}~\bibnamefont {Palummo}}, \bibinfo {author} {\bibfnamefont
  {A.}~\bibnamefont {Molina-S{\'{a}}nchez}}, \bibinfo {author} {\bibfnamefont
  {C.}~\bibnamefont {Hogan}}, \bibinfo {author} {\bibfnamefont
  {M.}~\bibnamefont {Grüning}}, \bibinfo {author} {\bibfnamefont
  {D.}~\bibnamefont {Varsano}}, \ and\ \bibinfo {author} {\bibfnamefont
  {A.}~\bibnamefont {Marini}},\ }\bibfield  {title} {\enquote {\bibinfo {title}
  {Many-body perturbation theory calculations using the \textsc{Yambo} code},}\
  }\href@noop {} {\bibfield  {journal} {\bibinfo  {journal} {J. Phys.: Condens.
  Matter}\ }\textbf {\bibinfo {volume} {31}},\ \bibinfo {pages} {325902}
  (\bibinfo {year} {2019})}\BibitemShut {NoStop}%
\bibitem [{\citenamefont {Schiferl}\ and\ \citenamefont
  {Barrett}(1969)}]{schiferl1969}%
  \BibitemOpen
  \bibfield  {author} {\bibinfo {author} {\bibfnamefont {D.}~\bibnamefont
  {Schiferl}}\ and\ \bibinfo {author} {\bibfnamefont {C.~S.}\ \bibnamefont
  {Barrett}},\ }\bibfield  {title} {\enquote {\bibinfo {title} {The crystal
  structure of arsenic at $4.2$, $78$ and $299$\si{\degree\kelvin}},}\
  }\href@noop {} {\bibfield  {journal} {\bibinfo  {journal} {J. Appl. Cryst.}\
  }\textbf {\bibinfo {volume} {2}},\ \bibinfo {pages} {30} (\bibinfo {year}
  {1969})}\BibitemShut {NoStop}%
\bibitem [{\citenamefont {König}\ \emph {et~al.}(2019)\citenamefont {König},
  \citenamefont {Fahy},\ and\ \citenamefont {Greer}}]{koenig2019}%
  \BibitemOpen
  \bibfield  {author} {\bibinfo {author} {\bibfnamefont {C.}~\bibnamefont
  {König}}, \bibinfo {author} {\bibfnamefont {S.}~\bibnamefont {Fahy}}, \ and\
  \bibinfo {author} {\bibfnamefont {J.~C.}\ \bibnamefont {Greer}},\ }\bibfield
  {title} {\enquote {\bibinfo {title} {Structural modification of thin
  {Bi}\hkl(1 1 1) films by passivation and native oxide model},}\ }\href@noop
  {} {\bibfield  {journal} {\bibinfo  {journal} {Phys. Rev. Materials}\
  }\textbf {\bibinfo {volume} {3}},\ \bibinfo {pages} {065002} (\bibinfo {year}
  {2019})}\BibitemShut {NoStop}%
\bibitem [{\citenamefont {Rozzi}\ \emph {et~al.}(2006)\citenamefont {Rozzi},
  \citenamefont {Varsano}, \citenamefont {Marini}, \citenamefont {Gross},\ and\
  \citenamefont {Rubio}}]{rozzi2006}%
  \BibitemOpen
  \bibfield  {author} {\bibinfo {author} {\bibfnamefont {C.~A.}\ \bibnamefont
  {Rozzi}}, \bibinfo {author} {\bibfnamefont {D.}~\bibnamefont {Varsano}},
  \bibinfo {author} {\bibfnamefont {A.}~\bibnamefont {Marini}}, \bibinfo
  {author} {\bibfnamefont {E.~K.~U.}\ \bibnamefont {Gross}}, \ and\ \bibinfo
  {author} {\bibfnamefont {A.}~\bibnamefont {Rubio}},\ }\bibfield  {title}
  {\enquote {\bibinfo {title} {Exact {Coulomb} cutoff technique for supercell
  calculations},}\ }\href@noop {} {\bibfield  {journal} {\bibinfo  {journal}
  {Phys. Rev. B}\ }\textbf {\bibinfo {volume} {73}},\ \bibinfo {pages} {205119}
  (\bibinfo {year} {2006})}\BibitemShut {NoStop}%
\bibitem [{\citenamefont {Madsen}\ and\ \citenamefont
  {Singh}(2006)}]{madsen2006}%
  \BibitemOpen
  \bibfield  {author} {\bibinfo {author} {\bibfnamefont {G.~K.~H.}\
  \bibnamefont {Madsen}}\ and\ \bibinfo {author} {\bibfnamefont {D.~J.}\
  \bibnamefont {Singh}},\ }\bibfield  {title} {\enquote {\bibinfo {title}
  {\textsc{BoltzTraP}. {A} code for calculating band-structure dependent
  quantities},}\ }\href@noop {} {\bibfield  {journal} {\bibinfo  {journal}
  {Comput. Phys. Commun.}\ }\textbf {\bibinfo {volume} {175}},\ \bibinfo
  {pages} {67} (\bibinfo {year} {2006})}\BibitemShut {NoStop}%
\bibitem [{\citenamefont {Madsen}\ \emph {et~al.}(2018)\citenamefont {Madsen},
  \citenamefont {Carrete},\ and\ \citenamefont {Verstraete}}]{madsen2018}%
  \BibitemOpen
  \bibfield  {author} {\bibinfo {author} {\bibfnamefont {G.~K.~H.}\
  \bibnamefont {Madsen}}, \bibinfo {author} {\bibfnamefont {J.}~\bibnamefont
  {Carrete}}, \ and\ \bibinfo {author} {\bibfnamefont {M.~J.}\ \bibnamefont
  {Verstraete}},\ }\bibfield  {title} {\enquote {\bibinfo {title}
  {\textsc{BoltzTraP2}, a program for interpolating band structures and
  calculating semi-classical transport coefficients},}\ }\href@noop {}
  {\bibfield  {journal} {\bibinfo  {journal} {Comput. Phys. Commun.}\ }\textbf
  {\bibinfo {volume} {231}},\ \bibinfo {pages} {140} (\bibinfo {year}
  {2018})}\BibitemShut {NoStop}%
\bibitem [{\citenamefont {Euwema}\ \emph {et~al.}(1969)\citenamefont {Euwema},
  \citenamefont {Stukel}, \citenamefont {Collins}, \citenamefont {DeWitt},\
  and\ \citenamefont {Shankland}}]{euwema1969}%
  \BibitemOpen
  \bibfield  {author} {\bibinfo {author} {\bibfnamefont {R.~N.}\ \bibnamefont
  {Euwema}}, \bibinfo {author} {\bibfnamefont {D.~J.}\ \bibnamefont {Stukel}},
  \bibinfo {author} {\bibfnamefont {T.~C.}\ \bibnamefont {Collins}}, \bibinfo
  {author} {\bibfnamefont {J.~S.}\ \bibnamefont {DeWitt}}, \ and\ \bibinfo
  {author} {\bibfnamefont {D.~G.}\ \bibnamefont {Shankland}},\ }\bibfield
  {title} {\enquote {\bibinfo {title} {Crystalline interpolation with
  applications to {Brillouin}-zone averages and energy-band interpolation},}\
  }\href@noop {} {\bibfield  {journal} {\bibinfo  {journal} {Phys. Rev.}\
  }\textbf {\bibinfo {volume} {178}},\ \bibinfo {pages} {1419} (\bibinfo {year}
  {1969})}\BibitemShut {NoStop}%
\bibitem [{\citenamefont {Shankland}(1971)}]{shankland1971}%
  \BibitemOpen
  \bibfield  {author} {\bibinfo {author} {\bibfnamefont {D.~G.}\ \bibnamefont
  {Shankland}},\ }\bibfield  {title} {\enquote {\bibinfo {title} {Fourier
  transformation by smooth interpolation},}\ }\href@noop {} {\bibfield
  {journal} {\bibinfo  {journal} {Int. J. Quantum Chem.}\ }\textbf {\bibinfo
  {volume} {5}},\ \bibinfo {pages} {497} (\bibinfo {year} {1971})}\BibitemShut
  {NoStop}%
\bibitem [{\citenamefont {Koelling}\ and\ \citenamefont
  {Wood}(1986)}]{koelling1986}%
  \BibitemOpen
  \bibfield  {author} {\bibinfo {author} {\bibfnamefont {D.~D.}\ \bibnamefont
  {Koelling}}\ and\ \bibinfo {author} {\bibfnamefont {J.~H.}\ \bibnamefont
  {Wood}},\ }\bibfield  {title} {\enquote {\bibinfo {title} {On the
  interpolation of eigenvalues and a resultant integration scheme},}\
  }\href@noop {} {\bibfield  {journal} {\bibinfo  {journal} {J. Comput. Phys}\
  }\textbf {\bibinfo {volume} {67}},\ \bibinfo {pages} {253} (\bibinfo {year}
  {1986})}\BibitemShut {NoStop}%
\bibitem [{\citenamefont {Pickett}\ \emph {et~al.}(1988)\citenamefont
  {Pickett}, \citenamefont {Krakauer},\ and\ \citenamefont
  {Allen}}]{pickett1988}%
  \BibitemOpen
  \bibfield  {author} {\bibinfo {author} {\bibfnamefont {W.~E.}\ \bibnamefont
  {Pickett}}, \bibinfo {author} {\bibfnamefont {H.}~\bibnamefont {Krakauer}}, \
  and\ \bibinfo {author} {\bibfnamefont {P.~B.}\ \bibnamefont {Allen}},\
  }\bibfield  {title} {\enquote {\bibinfo {title} {Smooth {Fourier}
  interpolation of periodic functions},}\ }\href@noop {} {\bibfield  {journal}
  {\bibinfo  {journal} {Phys. Rev. B}\ }\textbf {\bibinfo {volume} {38}},\
  \bibinfo {pages} {2721} (\bibinfo {year} {1988})}\BibitemShut {NoStop}%
\bibitem [{\citenamefont {Yao}\ \emph {et~al.}(2016)\citenamefont {Yao},
  \citenamefont {Zhu}, \citenamefont {Han}, \citenamefont {Guan}, \citenamefont
  {Liu}, \citenamefont {Qian},\ and\ \citenamefont {Jia}}]{yao2016}%
  \BibitemOpen
  \bibfield  {author} {\bibinfo {author} {\bibfnamefont {M.-Y.}\ \bibnamefont
  {Yao}}, \bibinfo {author} {\bibfnamefont {F.}~\bibnamefont {Zhu}}, \bibinfo
  {author} {\bibfnamefont {C.~Q.}\ \bibnamefont {Han}}, \bibinfo {author}
  {\bibfnamefont {D.~D.}\ \bibnamefont {Guan}}, \bibinfo {author}
  {\bibfnamefont {C.}~\bibnamefont {Liu}}, \bibinfo {author} {\bibfnamefont
  {D.}~\bibnamefont {Qian}}, \ and\ \bibinfo {author} {\bibfnamefont {J.-F.}\
  \bibnamefont {Jia}},\ }\bibfield  {title} {\enquote {\bibinfo {title}
  {Topologically nontrivial bismuth\hkl(1 1 1) thin films},}\ }\href@noop {}
  {\bibfield  {journal} {\bibinfo  {journal} {Sci. Rep.}\ }\textbf {\bibinfo
  {volume} {6}},\ \bibinfo {pages} {21326} (\bibinfo {year}
  {2016})}\BibitemShut {NoStop}%
\bibitem [{\citenamefont {Leontie}\ \emph {et~al.}(2000)\citenamefont
  {Leontie}, \citenamefont {Caraman},\ and\ \citenamefont
  {Rusu}}]{leontie2000}%
  \BibitemOpen
  \bibfield  {author} {\bibinfo {author} {\bibfnamefont {L.}~\bibnamefont
  {Leontie}}, \bibinfo {author} {\bibfnamefont {M.}~\bibnamefont {Caraman}}, \
  and\ \bibinfo {author} {\bibfnamefont {G.~I.}\ \bibnamefont {Rusu}},\
  }\bibfield  {title} {\enquote {\bibinfo {title} {On the photoconductivity of
  {Bi}$_2${O}$_3$ in thin films},}\ }\href@noop {} {\bibfield  {journal}
  {\bibinfo  {journal} {J. Optoelectron. Adv. M.}\ }\textbf {\bibinfo {volume}
  {2}},\ \bibinfo {pages} {385} (\bibinfo {year} {2000})}\BibitemShut {NoStop}%
\bibitem [{\citenamefont {Leontie}\ \emph {et~al.}(2001)\citenamefont
  {Leontie}, \citenamefont {Caraman}, \citenamefont {Delibaş},\ and\
  \citenamefont {Rusu}}]{leontie2001}%
  \BibitemOpen
  \bibfield  {author} {\bibinfo {author} {\bibfnamefont {L.}~\bibnamefont
  {Leontie}}, \bibinfo {author} {\bibfnamefont {M.}~\bibnamefont {Caraman}},
  \bibinfo {author} {\bibfnamefont {M.}~\bibnamefont {Delibaş}}, \ and\
  \bibinfo {author} {\bibfnamefont {G.~I.}\ \bibnamefont {Rusu}},\ }\bibfield
  {title} {\enquote {\bibinfo {title} {Optical properties of bismuth trioxide
  thin films},}\ }\href@noop {} {\bibfield  {journal} {\bibinfo  {journal}
  {Mater. Res. Bull.}\ }\textbf {\bibinfo {volume} {36}},\ \bibinfo {pages}
  {1629} (\bibinfo {year} {2001})}\BibitemShut {NoStop}%
\bibitem [{\citenamefont {Leontie}\ \emph {et~al.}(2002)\citenamefont
  {Leontie}, \citenamefont {Caraman}, \citenamefont {Alexe},\ and\
  \citenamefont {Harnagea}}]{leontie2002}%
  \BibitemOpen
  \bibfield  {author} {\bibinfo {author} {\bibfnamefont {L.}~\bibnamefont
  {Leontie}}, \bibinfo {author} {\bibfnamefont {M.}~\bibnamefont {Caraman}},
  \bibinfo {author} {\bibfnamefont {M.}~\bibnamefont {Alexe}}, \ and\ \bibinfo
  {author} {\bibfnamefont {C.}~\bibnamefont {Harnagea}},\ }\bibfield  {title}
  {\enquote {\bibinfo {title} {Structural and optical characteristics of
  bismuth oxide thin films},}\ }\href@noop {} {\bibfield  {journal} {\bibinfo
  {journal} {Surf. Sci.}\ }\textbf {\bibinfo {volume} {507-510}},\ \bibinfo
  {pages} {480} (\bibinfo {year} {2002})}\BibitemShut {NoStop}%
\bibitem [{\citenamefont {Walsh}\ \emph {et~al.}(2006)\citenamefont {Walsh},
  \citenamefont {Watson}, \citenamefont {Payne}, \citenamefont {Edgell},
  \citenamefont {Guo}, \citenamefont {Glans}, \citenamefont {Learmonth},\ and\
  \citenamefont {Smith}}]{walsh2006}%
  \BibitemOpen
  \bibfield  {author} {\bibinfo {author} {\bibfnamefont {A.}~\bibnamefont
  {Walsh}}, \bibinfo {author} {\bibfnamefont {G.~W.}\ \bibnamefont {Watson}},
  \bibinfo {author} {\bibfnamefont {D.~J.}\ \bibnamefont {Payne}}, \bibinfo
  {author} {\bibfnamefont {R.~G.}\ \bibnamefont {Edgell}}, \bibinfo {author}
  {\bibfnamefont {J.}~\bibnamefont {Guo}}, \bibinfo {author} {\bibfnamefont
  {P.-A.}\ \bibnamefont {Glans}}, \bibinfo {author} {\bibfnamefont
  {T.}~\bibnamefont {Learmonth}}, \ and\ \bibinfo {author} {\bibfnamefont
  {K.~E.}\ \bibnamefont {Smith}},\ }\bibfield  {title} {\enquote {\bibinfo
  {title} {Electronic structure of the $\alpha$ and $\delta$ phases of
  $\text{Bi}_2\text{O}_3$: A combined ab initio and x-ray spectroscopy
  study},}\ }\href@noop {} {\bibfield  {journal} {\bibinfo  {journal} {Phys.
  Rev. B}\ }\textbf {\bibinfo {volume} {73}},\ \bibinfo {pages} {235104}
  (\bibinfo {year} {2006})}\BibitemShut {NoStop}%
\bibitem [{\citenamefont {Gobrecht}\ \emph {et~al.}(1969)\citenamefont
  {Gobrecht}, \citenamefont {Seeck}, \citenamefont {Bergt}, \citenamefont
  {Märtens},\ and\ \citenamefont {Kossmann}}]{gobrecht1969}%
  \BibitemOpen
  \bibfield  {author} {\bibinfo {author} {\bibfnamefont {H.}~\bibnamefont
  {Gobrecht}}, \bibinfo {author} {\bibfnamefont {S.}~\bibnamefont {Seeck}},
  \bibinfo {author} {\bibfnamefont {H.-E.}\ \bibnamefont {Bergt}}, \bibinfo
  {author} {\bibfnamefont {A.}~\bibnamefont {Märtens}}, \ and\ \bibinfo
  {author} {\bibfnamefont {K.}~\bibnamefont {Kossmann}},\ }\bibfield  {title}
  {\enquote {\bibinfo {title} {{Über Untersuchungen an
  Wismutoxid-Aufdampfschichten I. Herstellung sowie elektrische und optische
  Eigenschaften}},}\ }\href@noop {} {\bibfield  {journal} {\bibinfo  {journal}
  {Phys. Stat. Sol.}\ }\textbf {\bibinfo {volume} {33}},\ \bibinfo {pages}
  {599} (\bibinfo {year} {1969})}\BibitemShut {NoStop}%
\bibitem [{Note1()}]{Note1}%
  \BibitemOpen
  \bibinfo {note} {The band gap is wide enough so that the coupling of the
  oxide to the Bi states at the Fermi level is small.}\BibitemShut {Stop}%
\bibitem [{\citenamefont {König}\ \emph {et~al.}(2021)\citenamefont {König},
  \citenamefont {Greer},\ and\ \citenamefont {Fahy}}]{koenig2021a}%
  \BibitemOpen
  \bibfield  {author} {\bibinfo {author} {\bibfnamefont {C.}~\bibnamefont
  {König}}, \bibinfo {author} {\bibfnamefont {J.~C.}\ \bibnamefont {Greer}}, \
  and\ \bibinfo {author} {\bibfnamefont {S.}~\bibnamefont {Fahy}},\ }\bibfield
  {title} {\enquote {\bibinfo {title} {Effect of strain and many-body
  corrections on the band inversions and topology of bismuth},}\ }\href@noop {}
  {\bibfield  {journal} {\bibinfo  {journal} {Phys. Rev. B}\ }\textbf {\bibinfo
  {volume} {104}},\ \bibinfo {pages} {035127} (\bibinfo {year}
  {2021})}\BibitemShut {NoStop}%
\bibitem [{Note2()}]{Note2}%
  \BibitemOpen
  \bibinfo {note} {At the same time the material is pushed towards the
  topologically non-trivial state (decreasing L gap).}\BibitemShut {Stop}%
\bibitem [{\citenamefont {Jin}\ \emph {et~al.}(2020)\citenamefont {Jin},
  \citenamefont {Yeom},\ and\ \citenamefont {Liu}}]{jin2020}%
  \BibitemOpen
  \bibfield  {author} {\bibinfo {author} {\bibfnamefont {K.-H.}\ \bibnamefont
  {Jin}}, \bibinfo {author} {\bibfnamefont {H.~W.}\ \bibnamefont {Yeom}}, \
  and\ \bibinfo {author} {\bibfnamefont {F.}~\bibnamefont {Liu}},\ }\bibfield
  {title} {\enquote {\bibinfo {title} {Doping-induced topological phase
  transition in {Bi}: The role of quantum electronic stress},}\ }\href@noop {}
  {\bibfield  {journal} {\bibinfo  {journal} {Phys. Rev. B}\ }\textbf {\bibinfo
  {volume} {101}},\ \bibinfo {pages} {035111} (\bibinfo {year}
  {2020})}\BibitemShut {NoStop}%
\bibitem [{\citenamefont {Ito}\ \emph {et~al.}(2020)\citenamefont {Ito},
  \citenamefont {Arita}, \citenamefont {Haruyama}, \citenamefont {Feng},
  \citenamefont {Chen}, \citenamefont {Namatame}, \citenamefont {Taniguchi},
  \citenamefont {Cheng}, \citenamefont {Bian}, \citenamefont {Tang},
  \citenamefont {Chiang}, \citenamefont {Sugino}, \citenamefont {Komori},\ and\
  \citenamefont {Matsuda}}]{ito2020}%
  \BibitemOpen
  \bibfield  {author} {\bibinfo {author} {\bibfnamefont {S.}~\bibnamefont
  {Ito}}, \bibinfo {author} {\bibfnamefont {M.}~\bibnamefont {Arita}}, \bibinfo
  {author} {\bibfnamefont {J.}~\bibnamefont {Haruyama}}, \bibinfo {author}
  {\bibfnamefont {B.}~\bibnamefont {Feng}}, \bibinfo {author} {\bibfnamefont
  {W.-C.}\ \bibnamefont {Chen}}, \bibinfo {author} {\bibfnamefont
  {H.}~\bibnamefont {Namatame}}, \bibinfo {author} {\bibfnamefont
  {M.}~\bibnamefont {Taniguchi}}, \bibinfo {author} {\bibfnamefont {C.-M.}\
  \bibnamefont {Cheng}}, \bibinfo {author} {\bibfnamefont {G.}~\bibnamefont
  {Bian}}, \bibinfo {author} {\bibfnamefont {S.-J.}\ \bibnamefont {Tang}},
  \bibinfo {author} {\bibfnamefont {T.-C.}\ \bibnamefont {Chiang}}, \bibinfo
  {author} {\bibfnamefont {O.}~\bibnamefont {Sugino}}, \bibinfo {author}
  {\bibfnamefont {F.}~\bibnamefont {Komori}}, \ and\ \bibinfo {author}
  {\bibfnamefont {I.}~\bibnamefont {Matsuda}},\ }\bibfield  {title} {\enquote
  {\bibinfo {title} {Surface-state {Coulomb} repulsion accelerates a
  metal-insulator transition in topological semimetal nanofilms},}\ }\href@noop
  {} {\bibfield  {journal} {\bibinfo  {journal} {Sci. Adv.}\ }\textbf {\bibinfo
  {volume} {6}},\ \bibinfo {pages} {eaaz5015} (\bibinfo {year}
  {2020})}\BibitemShut {NoStop}%
\bibitem [{\citenamefont {Nagao}\ \emph {et~al.}(2004)\citenamefont {Nagao},
  \citenamefont {Sadowski}, \citenamefont {Saito}, \citenamefont {Yaginuma},
  \citenamefont {Fujikawa}, \citenamefont {Kogure}, \citenamefont {Ohno},
  \citenamefont {Hasegawa}, \citenamefont {Hasegawa},\ and\ \citenamefont
  {Sakurai}}]{nagao2004}%
  \BibitemOpen
  \bibfield  {author} {\bibinfo {author} {\bibfnamefont {T.}~\bibnamefont
  {Nagao}}, \bibinfo {author} {\bibfnamefont {J.~T.}\ \bibnamefont {Sadowski}},
  \bibinfo {author} {\bibfnamefont {M.}~\bibnamefont {Saito}}, \bibinfo
  {author} {\bibfnamefont {S.}~\bibnamefont {Yaginuma}}, \bibinfo {author}
  {\bibfnamefont {Y.}~\bibnamefont {Fujikawa}}, \bibinfo {author}
  {\bibfnamefont {T.}~\bibnamefont {Kogure}}, \bibinfo {author} {\bibfnamefont
  {T.}~\bibnamefont {Ohno}}, \bibinfo {author} {\bibfnamefont {Y.}~\bibnamefont
  {Hasegawa}}, \bibinfo {author} {\bibfnamefont {S.}~\bibnamefont {Hasegawa}},
  \ and\ \bibinfo {author} {\bibfnamefont {T.}~\bibnamefont {Sakurai}},\
  }\bibfield  {title} {\enquote {\bibinfo {title} {Nanofilm allotrope and phase
  transformation of ultrathin {Bi} film on {Si}(111)-7$\times$7},}\ }\href@noop
  {} {\bibfield  {journal} {\bibinfo  {journal} {Phys. Rev. Lett.}\ }\textbf
  {\bibinfo {volume} {93}},\ \bibinfo {pages} {105501} (\bibinfo {year}
  {2004})}\BibitemShut {NoStop}%
\bibitem [{\citenamefont {Kammler}\ and\ \citenamefont {Horn-von
  Hoegen}(2005)}]{kammler2005}%
  \BibitemOpen
  \bibfield  {author} {\bibinfo {author} {\bibfnamefont {M.}~\bibnamefont
  {Kammler}}\ and\ \bibinfo {author} {\bibfnamefont {M.}~\bibnamefont {Horn-von
  Hoegen}},\ }\bibfield  {title} {\enquote {\bibinfo {title} {Low energy
  electron diffraction of epitaxial growth of bismuth on {Si}\hkl(1 1 1)},}\
  }\href@noop {} {\bibfield  {journal} {\bibinfo  {journal} {Surf. Sci.}\
  }\textbf {\bibinfo {volume} {576}},\ \bibinfo {pages} {56} (\bibinfo {year}
  {2005})}\BibitemShut {NoStop}%
\bibitem [{\citenamefont {Nagao}\ \emph {et~al.}(2005)\citenamefont {Nagao},
  \citenamefont {Yaginuma}, \citenamefont {Saito}, \citenamefont {Kogure},
  \citenamefont {Sadowski}, \citenamefont {Ohno}, \citenamefont {Hasegawa},\
  and\ \citenamefont {Sakurai}}]{nagao2005}%
  \BibitemOpen
  \bibfield  {author} {\bibinfo {author} {\bibfnamefont {T.}~\bibnamefont
  {Nagao}}, \bibinfo {author} {\bibfnamefont {S.}~\bibnamefont {Yaginuma}},
  \bibinfo {author} {\bibfnamefont {M.}~\bibnamefont {Saito}}, \bibinfo
  {author} {\bibfnamefont {T.}~\bibnamefont {Kogure}}, \bibinfo {author}
  {\bibfnamefont {J.~T.}\ \bibnamefont {Sadowski}}, \bibinfo {author}
  {\bibfnamefont {T.}~\bibnamefont {Ohno}}, \bibinfo {author} {\bibfnamefont
  {S.}~\bibnamefont {Hasegawa}}, \ and\ \bibinfo {author} {\bibfnamefont
  {T.}~\bibnamefont {Sakurai}},\ }\bibfield  {title} {\enquote {\bibinfo
  {title} {Strong lateral growth and crystallization via two-dimensional
  allotropic transformation of semi-metal {Bi} film},}\ }\href@noop {}
  {\bibfield  {journal} {\bibinfo  {journal} {Surf. Sci.}\ }\textbf {\bibinfo
  {volume} {590}},\ \bibinfo {pages} {247} (\bibinfo {year}
  {2005})}\BibitemShut {NoStop}%
\bibitem [{\citenamefont {Yaginuma}\ \emph {et~al.}(2007)\citenamefont
  {Yaginuma}, \citenamefont {Nagao}, \citenamefont {Sadowski}, \citenamefont
  {Saito}, \citenamefont {Nagaoka}, \citenamefont {Fujikawa}, \citenamefont
  {Sakurai},\ and\ \citenamefont {Nakayama}}]{yaginuma2007}%
  \BibitemOpen
  \bibfield  {author} {\bibinfo {author} {\bibfnamefont {S.}~\bibnamefont
  {Yaginuma}}, \bibinfo {author} {\bibfnamefont {T.}~\bibnamefont {Nagao}},
  \bibinfo {author} {\bibfnamefont {J.~T.}\ \bibnamefont {Sadowski}}, \bibinfo
  {author} {\bibfnamefont {M.}~\bibnamefont {Saito}}, \bibinfo {author}
  {\bibfnamefont {K.}~\bibnamefont {Nagaoka}}, \bibinfo {author} {\bibfnamefont
  {Y.}~\bibnamefont {Fujikawa}}, \bibinfo {author} {\bibfnamefont
  {T.}~\bibnamefont {Sakurai}}, \ and\ \bibinfo {author} {\bibfnamefont
  {T.}~\bibnamefont {Nakayama}},\ }\bibfield  {title} {\enquote {\bibinfo
  {title} {Origin of flat morphology and high crystallinity of ultrathin
  bismuth films},}\ }\href@noop {} {\bibfield  {journal} {\bibinfo  {journal}
  {Surf. Sci.}\ }\textbf {\bibinfo {volume} {601}},\ \bibinfo {pages} {3593}
  (\bibinfo {year} {2007})}\BibitemShut {NoStop}%
\bibitem [{Note3()}]{Note3}%
  \BibitemOpen
  \bibinfo {note} {Only in the artificial situation where hydrogen is attached
  to an otherwise unrelaxed slab, the surface states are sufficiently deformed
  to open a small band gap, but not entirely removed.}\BibitemShut {Stop}%
\bibitem [{sup()}]{supplemental}%
  \BibitemOpen
  \href@noop {} {}\bibinfo {note} {See Supplemental Material at URL for the
  effect of the many-body interaction corrections in the context of a scissors
  operator.}\BibitemShut {Stop}%
\bibitem [{\citenamefont {Qiu}\ \emph {et~al.}(2017)\citenamefont {Qiu},
  \citenamefont {da~Jornada},\ and\ \citenamefont {Louie}}]{qiu2017}%
  \BibitemOpen
  \bibfield  {author} {\bibinfo {author} {\bibfnamefont {D.~Y.}\ \bibnamefont
  {Qiu}}, \bibinfo {author} {\bibfnamefont {F.~H.}\ \bibnamefont {da~Jornada}},
  \ and\ \bibinfo {author} {\bibfnamefont {S.~G.}\ \bibnamefont {Louie}},\
  }\bibfield  {title} {\enquote {\bibinfo {title} {Environmental screening
  effects in 2d materials: Renormalization of the bandgap, electronic
  structure, and optical spectra of few-layer black phosphorus},}\ }\href@noop
  {} {\bibfield  {journal} {\bibinfo  {journal} {Nano Lett.}\ }\textbf
  {\bibinfo {volume} {17}},\ \bibinfo {pages} {4706} (\bibinfo {year}
  {2017})}\BibitemShut {NoStop}%
\bibitem [{\citenamefont {Du}\ \emph {et~al.}(2016)\citenamefont {Du},
  \citenamefont {Sun}, \citenamefont {Liu}, \citenamefont {Wu}, \citenamefont
  {Wang}, \citenamefont {Tian}, \citenamefont {Zhao}, \citenamefont {Luo},
  \citenamefont {Yang}, \citenamefont {Wang},\ and\ \citenamefont
  {Hou}}]{du2016}%
  \BibitemOpen
  \bibfield  {author} {\bibinfo {author} {\bibfnamefont {H.}~\bibnamefont
  {Du}}, \bibinfo {author} {\bibfnamefont {X.}~\bibnamefont {Sun}}, \bibinfo
  {author} {\bibfnamefont {X.}~\bibnamefont {Liu}}, \bibinfo {author}
  {\bibfnamefont {X.}~\bibnamefont {Wu}}, \bibinfo {author} {\bibfnamefont
  {J.}~\bibnamefont {Wang}}, \bibinfo {author} {\bibfnamefont {M.}~\bibnamefont
  {Tian}}, \bibinfo {author} {\bibfnamefont {A.}~\bibnamefont {Zhao}}, \bibinfo
  {author} {\bibfnamefont {Y.}~\bibnamefont {Luo}}, \bibinfo {author}
  {\bibfnamefont {J.}~\bibnamefont {Yang}}, \bibinfo {author} {\bibfnamefont
  {B.}~\bibnamefont {Wang}}, \ and\ \bibinfo {author} {\bibfnamefont {J.~G.}\
  \bibnamefont {Hou}},\ }\bibfield  {title} {\enquote {\bibinfo {title}
  {Surface {Landau} levels and spin states in bismuth \hkl(1 1 1) ultrathin
  films},}\ }\href@noop {} {\bibfield  {journal} {\bibinfo  {journal} {Nat.
  Commun.}\ }\textbf {\bibinfo {volume} {7}},\ \bibinfo {pages} {10814}
  (\bibinfo {year} {2016})}\BibitemShut {NoStop}%
\bibitem [{\citenamefont {O’Mahony}\ \emph {et~al.}(2019)\citenamefont
  {O’Mahony}, \citenamefont {Murphy-Armando}, \citenamefont {Murray},
  \citenamefont {Querales-Flores}, \citenamefont {Savić},\ and\ \citenamefont
  {Fahy}}]{omahony2019}%
  \BibitemOpen
  \bibfield  {author} {\bibinfo {author} {\bibfnamefont {S.~M.}\ \bibnamefont
  {O’Mahony}}, \bibinfo {author} {\bibfnamefont {F.}~\bibnamefont
  {Murphy-Armando}}, \bibinfo {author} {\bibfnamefont {{\'E}.~D.}\ \bibnamefont
  {Murray}}, \bibinfo {author} {\bibfnamefont {J.~D.}\ \bibnamefont
  {Querales-Flores}}, \bibinfo {author} {\bibfnamefont {I.}~\bibnamefont
  {Savić}}, \ and\ \bibinfo {author} {\bibfnamefont {S.}~\bibnamefont
  {Fahy}},\ }\bibfield  {title} {\enquote {\bibinfo {title} {Ultrafast
  relaxation of symmetry-breaking photo-induced atomic forces},}\ }\href@noop
  {} {\bibfield  {journal} {\bibinfo  {journal} {Phys. Rev. Lett.}\ }\textbf
  {\bibinfo {volume} {123}},\ \bibinfo {pages} {087401} (\bibinfo {year}
  {2019})}\BibitemShut {NoStop}%
\bibitem [{\citenamefont {Liu}\ \emph {et~al.}(2011)\citenamefont {Liu},
  \citenamefont {Liu}, \citenamefont {Wu}, \citenamefont {Duan}, \citenamefont
  {Liu},\ and\ \citenamefont {Wu}}]{liu2011}%
  \BibitemOpen
  \bibfield  {author} {\bibinfo {author} {\bibfnamefont {Z.}~\bibnamefont
  {Liu}}, \bibinfo {author} {\bibfnamefont {C.-X.}\ \bibnamefont {Liu}},
  \bibinfo {author} {\bibfnamefont {Y.-S.}\ \bibnamefont {Wu}}, \bibinfo
  {author} {\bibfnamefont {W.-H.}\ \bibnamefont {Duan}}, \bibinfo {author}
  {\bibfnamefont {F.}~\bibnamefont {Liu}}, \ and\ \bibinfo {author}
  {\bibfnamefont {J.}~\bibnamefont {Wu}},\ }\bibfield  {title} {\enquote
  {\bibinfo {title} {Stable nontrivial {Z}$_2$ topology in ultrathin
  {Bi}\hkl(111) films: {A} first-principles study},}\ }\href@noop {} {\bibfield
   {journal} {\bibinfo  {journal} {Phys. Rev. Lett.}\ }\textbf {\bibinfo
  {volume} {107}},\ \bibinfo {pages} {136805} (\bibinfo {year}
  {2011})}\BibitemShut {NoStop}%
\bibitem [{\citenamefont {Momma}\ and\ \citenamefont
  {Izumi}(2011)}]{momma2011}%
  \BibitemOpen
  \bibfield  {author} {\bibinfo {author} {\bibfnamefont {K.}~\bibnamefont
  {Momma}}\ and\ \bibinfo {author} {\bibfnamefont {F.}~\bibnamefont {Izumi}},\
  }\bibfield  {title} {\enquote {\bibinfo {title} {\textsc{Vesta} 3 for
  three-dimensional visualization of crystal, volumetric and morphology
  data},}\ }\href@noop {} {\bibfield  {journal} {\bibinfo  {journal} {J. Appl.
  Crystallogr.}\ }\textbf {\bibinfo {volume} {44}},\ \bibinfo {pages} {1272}
  (\bibinfo {year} {2011})}\BibitemShut {NoStop}%
\end{thebibliography}%

\end{document}


\title{\textit{Supplemental Material:} Electronic properties of bismuth nanostructures}

\author{Christian \surname{König}}
\affiliation{Tyndall National Institute, University College Cork, Lee Maltings, Cork T12 R5CP, Ireland}
\author{James C. \surname{Greer}}
\affiliation{Nottingham Ningbo New Material Institute and Department of Electrical and Electronic Engineering, University of Nottingham Ningbo China, 199 Taikang East Road, Ningbo, 315100, China}
\author{Stephen \surname{Fahy}}
\affiliation{Tyndall National Institute, University College Cork, Lee Maltings, Cork T12 R5CP, Ireland}
\affiliation{Department of Physics, University College Cork, College Road, Cork T12 K8AF, Ireland}

\date{\today}

\begin{abstract}
In this supplemental we show some more details regarding the quasiparticle corrections
in the thin films. It is often assumed that the G$_0$W$_0$ corrected band structure
can be approximated with a DFT band structure, where the conduction bands are shifted
uniformly to higher energies (scissors operator).
Here, the corrections resemble a step-like function of the DFT energy or band index
quite well, in particular for the five bilayer thick slab. However, the states at
$\overline{\Gamma}$ do not shift to higher energies by the same amount which smears out the sharp
transition from the valence to the conduction bands.
\end{abstract}

\maketitle

\begin{figure*} 
  \includegraphics{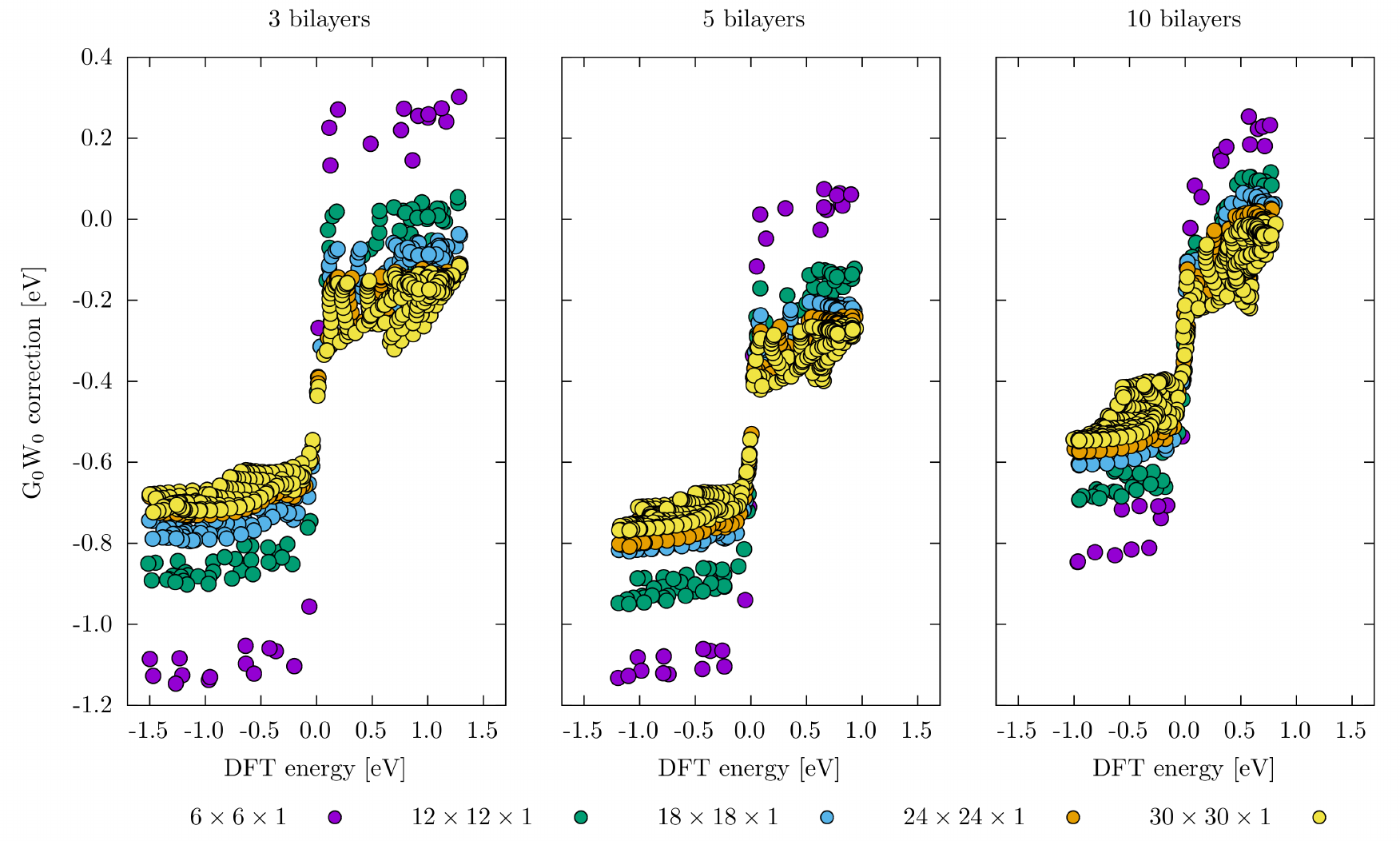}
  \caption{
  G$_0$W$_0$ corrections from the slab calculations as a function of the DFT energy. The results are shown for up to $30 \times 30$ $\mathbf{k}$-points.
}
\end{figure*}

\begin{figure*} 
  \center
  \includegraphics{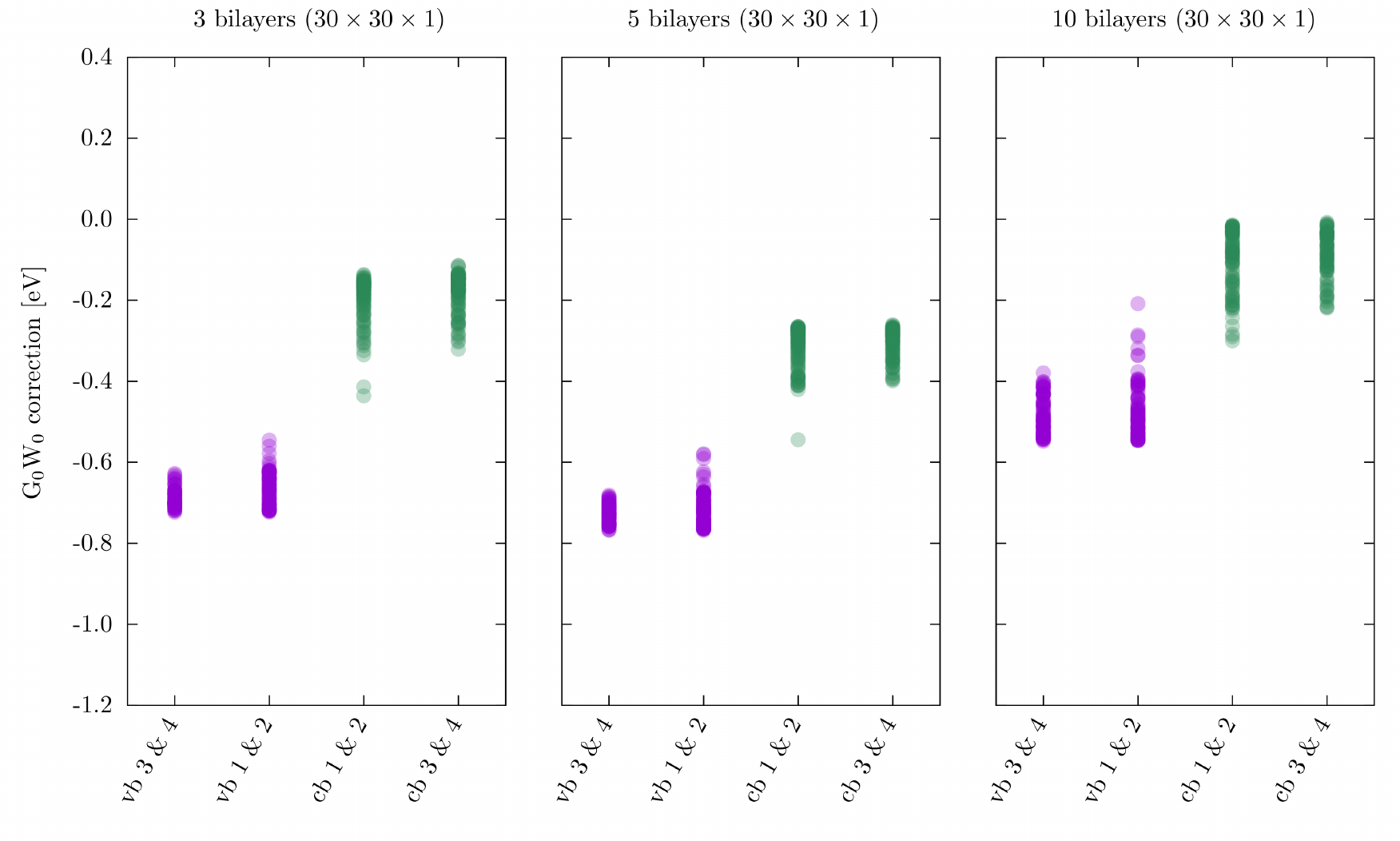}
  \caption{
  G$_0$W$_0$ corrections from the slab calculations as a function of the band index (using $30$ $\mathbf{k}$-points in each periodic direction). The plot shows that despite the variations within each band, there is a clear offset between the valence and conduction bands.
}
\end{figure*}